\documentclass[10pt,journal,compsoc]{IEEEtran}
\IEEEoverridecommandlockouts
\usepackage[nocompress]{cite}
\usepackage{cite}
\usepackage{amsmath,amssymb,amsfonts}
\usepackage{paralist}
\usepackage{algorithmic}
\usepackage{graphicx}
\usepackage{textcomp}
\usepackage{multirow}
\usepackage{subfigure}
\usepackage{url}
\usepackage{acronym}
\usepackage[english]{babel}
\usepackage{soul,color}
\usepackage[table,xcdraw]{xcolor}
\usepackage{xcolor}
\usepackage{multirow}
\usepackage[normalem]{ulem}

\def\BibTeX{{\rm B\kern-.05em{\sc i\kern-.025em b}\kern-.08em
    T\kern-.1667em\lower.7ex\hbox{E}\kern-.125emX}}

\definecolor{Yellow}{rgb}{1,0.9,0.7}
\definecolor{Pink}{rgb}{1,0.85,0.85}
\definecolor{AntiqueWhite}{rgb}{0.9,0.9,0.9}

\newcommand{\NOTE}[1]%
{
\noindent
\fboxsep=2mm\fcolorbox{black}{AntiqueWhite}{\parbox{0.95\columnwidth}
{\textbf{NOTE: } #1}
}
}

\begin{document}

\acrodef{ABFT}[ABFT]{Algorithm-Based Fault Tolerance}
\acrodef{ADAS}[ADAS]{Advanced Driver Assistance System}
\acrodef{API}[API]{Application Programming Interface}
\acrodef{CNN}[CNN]{Convolutional Neural Network}
\acrodef{CV}[CV]{Computer Vision}
\acrodef{DSP}[DSP]{Digital Signal Processing}
\acrodef{DUT}[DUT]{Design Under Test}
\acrodef{DWC}[DWC]{Duplication with Comparison}
\acrodef{ECC}[ECC]{Error Correction Code}
\acrodef{FI}[FI]{Fault Injection}
\acrodef{FES}[FES]{Fuctional Error Simulation}
\acrodef{GEMM}[GEMM]{General Matrix Multiplications}
\acrodef{GPU}[GPU]{Graphic Processing Unit}
\acrodef{HDL}[HDL]{Hardware Description Language}
\acrodef{IP}[IP]{Intellectual Property}
\acrodef{ISA}[ISA]{Instruction Set Architecture}
\acrodef{ML}[ML]{Machine Learning}
\acrodef{QOS}[QoS]{Quality-of-Service}
\acrodef{RELU}[ReLU]{Rectified Linear Unit}
\acrodef{RTL}[RTL]{Register-Transfer Level}
\acrodef{SAE}[SAE]{Society of Automotive Engineers}
\acrodef{SDC}[SDC]{Silent Data Corruption}
\acrodef{SEU}[SEU]{Single Event Upset}
\acrodef{TMR}[TMR]{Triple Modular Redundancy}
\acrodef{TRC}[TRC]{Two-Rail Checker}
\acrodef{SIMD}[SIMD]{Single Instruction Multiple Data}
\acrodef{SIMT}[SIMT]{Single Instruction Multiple Thread}
\acrodef{SM}[SM]{Streaming Multiprocessor}
\acrodef{AVF}[AVF]{Architectural Vulnerability Factor}

%\acrodef{ourtool}[the proposed approach]{the proposed approach}
%\acrodef{Ourtool}[The proposed approach]{The proposed approach}
%\acrodef{ourtool}[FILTER]{\textbf{F}unct\textbf{I}onal error simu\textbf{L}a\textbf{T}ion of soft \textbf{ER}ror effects in \acp{CNN}}
%\acrodef{Ourtool}[FILTER]{FILTER}

\acrodef{ourtool}[CLASSES]{\textbf{C}ross-\textbf{L}ayer \textbf{A}nalysi\textbf{S} framework for \textbf{S}oft-\textbf{E}rrors effect\textbf{S} in CNNs}

%% acronimo visualizzato al contrario, prima compresso e poi espanso
\newcommand{\acr}[1]{\acs{#1} (\aclu{#1})}
%FILTERED: FunctIonal error simuLaTion of soft ERror effects in XLs

%\title{Accurate and Fast Reliability Evaluation \\ of Convolutional Neural Networks \\ through Cross-Layer Analysis}
%\title{Cross-Layer Reliability Analysis of Convolutional Neural Networks accelerated onto GPUs}
%\title{Cross-Layer Reliability Analysis \\ of Convolutional Neural Networks}
%\title{Accurate and Fast Error Simulation \\of Convolutional Neural Networks \\accelerated onto GPUs}
\title{Fast and Accurate Error Simulation for CNNs against Soft Errors}
%\title{Accurate and Fast Error Simulation of Convolutional Neural Networks}

\author{Cristiana~Bolchini,~\IEEEmembership{Senior Member,~IEEE,}
        Luca~Cassano,~\IEEEmembership{Member,~IEEE,}\\
        Antonio~Miele,~\IEEEmembership{Senior Member,~IEEE,}
        Alessandro~Toschi
\IEEEcompsocitemizethanks{\IEEEcompsocthanksitem Authors are with the Dipartimento di Elettronica, Informazione e Bioingegneria, Politecnico di Milano, Italy.\protect\\
E-mail: \{first\_name.last\_name\}@polimi.it}% <-this % stops an unwanted space
}

\IEEEtitleabstractindextext{%

\acresetall

\begin{abstract}
The great quest for adopting AI-based computation for safety-/mission-critical applications motivates the interest towards methods for assessing the robustness of the application w.r.t. not only its training/tuning but also errors due to faults, in particular soft errors, affecting the underlying hardware. Two strategies exist: architecture-level fault injection and application-level functional error simulation. 
We present a framework for the reliability analysis of \acp{CNN} via an error simulation engine that exploits a set of validated error models extracted from a detailed fault injection campaign.
These error models are defined based on the corruption patterns of the output of the \ac{CNN} operators induced by faults and bridge the gap between fault injection and error simulation, exploiting the advantages of both approaches.
We compared our methodology against SASSIFI for the accuracy of functional error simulation w.r.t. fault injection, and against TensorFI in terms of speedup for the error simulation strategy. Experimental results show that our methodology achieves about 99\% accuracy of the fault effects w.r.t. SASSIFI, and a speedup ranging from 44x up to 63x w.r.t. TensorFI, that only implements a limited set of error models.
\end{abstract}

\begin{IEEEkeywords}
Soft Errors, Convolutional Neural Networks, Cross-layer Reliability Analysis, Error Modeling and Simulation, Fault Injection
\end{IEEEkeywords}}

\maketitle

\acresetall

\section{Introduction}
There is a growing interest in employing \acp{CNN} for perception functionalities in a wide range of application domains, including safety- and mission-critical ones (e.g., automotive, robots and avionics and aerospace). As a representative example, let us consider the \ac{ADAS} in the automotive scenario~\cite{campbell2010autonomous}; 
\acp{CNN} are employed to detect the lanes of the track, to identify pedestrians and obstacles, to interpret road signs and traffic lights~\cite{8814260, 8611202}. Based on such observations, subsequent planning modules in the \ac{ADAS} take trajectory and control decisions. 
In this context, \acp{CNN} are generally executed on \acp{GPU} since the \ac{SIMD} architecture is particularly well-suited to speed up the highly data-parallel elaborations that characterize these applications, allowing to meet the strict real-time requirements imposed by the \ac{ADAS}~\cite{8814260}. 

The design of digital systems in the automotive domain is regulated by the ISO~26262 standard~\cite{iso26262}, and the functionalities offered by an \ac{ADAS} are classified and regulated by the \ac{SAE}. Both standards require the system to expose a very high degree of reliability and to provide fault detection/management mechanisms. 
Although radiation-induced faults have historically been considered a concern mainly in the aerospace domain, it has been demonstrated that soft errors, such as \acp{SEU}~\cite{KH2004}, may interfere with the functionality of electronic systems also at the ground-level~\cite{normand1996single}, with an estimated ratio of two transient faults every thousand billion hours, on average. In 2019 the number of cars traveling in Europe has been 268 millions (see~\cite{ReportVe19:online}), leading to an estimate of a fault per car every 3.7 hours, which may be a concern.

It is thus paramount to be able to evaluate the resiliency of \ac{CNN}-based applications against soft errors within their application context, to determine how to harden the system to achieve the desired/required reliability level.
\acp{CNN}, and in general image processing and \ac{ML} applications, may expose an intrinsic degree of fault resilience due to several reasons: 
\begin{inparaenum}[i)] 
\item they may deal with noisy inputs (e.g., sensors) or data quantizations, 
\item their outputs may be probabilistic estimates, or 
\item produced data (such as image) may be used by a human, whose perceptual limitations provide resiliency to a certain level of inexactness~\cite{Mit2016}. 
\end{inparaenum}
Furthermore, studies have investigated how the internal redundancies of \ac{ML} models offer a certain degree of fault resilience~\cite{rech2019b,ibrahim2020,HBL2020,BINFI}.

To define novel hardening techniques specific for \acp{CNN}, it is necessary to be able to accurately identify the vulnerabilities against faults of the application and of the various parts of it; that is to analyze the effects of the faults occurring in the underlying hardware platform on the behavior of the application itself. 
Thus, for this class of applications, hardening techniques are moving from the classical bit-wise correct/corrupted checking of the outputs towards a \textit{usability-based} classification, to answer the question ``is the downstream system able to correctly carry out its task with the produced, possibly corrupted, output?''~\cite{BBC+2020}.

\textbf{To support this strategy, the novel contribution we propose is an accurate and fast cross-layer framework for the reliability analysis of \ac{CNN}-based applications against soft errors, that studies the effects of such faults not merely on the different outputs, but also on the final functionality within the entire system.}

\acused{FI}
\acused{FES}

If we exclude radiation testing, which is very expensive and does not provide enough internal controllability and observability, most relevant methods for reliability analysis fall into two main families: architecture-level \acl{FI}~\cite{fang2014gpu,sassifi,li2016understanding} and application-level \acl{FES}~\cite{chen2020,LSS+2017,bosio2019}. \textit{Architecture-Level \acl{FI}}, \acs{FI} in short, provides controllability and observability, as well as high accuracy, because of the ability of emulating the physical effects of the faults in the architecture components. The main drawback of \ac{FI} techniques is the long time and effort required to integrate the \ac{FI} features within the application under analysis and then to deploy them on the target architecture. Moreover, since faults may either be not activated or produce no observable error, extensive \ac{FI} campaigns may be required to collect a statistically relevant amount of corrupted data. \textit{Application-Level \acl{FES}}, \ac{FES} in short, provides shorter development and deployment times, it can be applied very early during the design process, and it ensures that every experiment produces a corrupted output. The main drawback resides in the accuracy; it is crucial that the adopted \textit{error models} are representative of the effects that faults in the hardware may induce in the data processed by the application. Indeed, as discussed in~\cite{PG2021}, \ac{FES} tools frequently used in past works adopt inaccurate error models; therefore, there is the necessity to fill the gap between the fault occurrence at the hardware layer and the error manifestation at software one.

This paper proposes \acr{ourtool}, a novel cross-layer framework for an early, accurate and fast reliability analysis of \acp{CNN} accelerated onto \acp{GPU} when affected by \acp{SEU}. The proposal bridges the gap between fault injection and error simulation, by leveraging both methods in different phases exploiting their advantages. 
\ac{ourtool} consists of two parts. The former is a systematic approach for accurate error modeling:
\begin{inparaenum}[i)]
\item an architecture-level \ac{FI} tool is exploited to perform an extensive fault injection campaign on each one of the basic operators used in \acp{CNN}, collecting all corrupted data;
\item the outcomes are then automatically analyzed to derive the corruption patterns that faults may induce in the output of each considered operators; 
\item corruption patterns are then used to define a set of error models integrated into the error simulation engine, constituting the second part of the framework.
\end{inparaenum}
In particular, for the \ac{FES} part we exploited TensorFlow~\cite{tensorflow2015-whitepaper}, at present one of the most popular \ac{ML} design frameworks. 
Error simulator is used to analyze in a faster and easier way the reliability of the entire application. Moreover, according to the peculiarities of the considered class of applications, the error simulator analyses \ac{CNN} outputs by means of a usability-based classification strategy, specifically defined by the designer.
As a result, our proposal enables an early but accurate evaluation of the reliability of any \ac{CNN}.
The contributions of our proposal can be summarized as follows: 
\begin{itemize}
\item a cross-layer framework for evaluating \acp{CNN} robustness against fault affecting the underlying hardware platform;
\item a designed and implemented semi-automated error modeling tool based on architecture-level \ac{FI} in \ac{GPU};
\item a set of validated functional error models for \acp{CNN} executed onto \ac{GPU}; and
\item a designed and implemented fully-automated error simulator integrated within TensorFlow.
\end{itemize}

The solution we designed is flexible, allowing the framework to integrate a different \ac{FI} tool (for instance the recent~\cite{rech2019, NVBITFI} and new ones that will emerge), based on the adopted development environment, to produce the error models required to support the \ac{FES} appropriately.

We analyze the framework performance and outcomes by comparing it against two representative \ac{FI} and \ac{FES} engines for \acp{CNN} accelerated onto \ac{GPU}, namely SASSIFI~\cite{sassifi} and TensorFI~\cite{chen2020}. Due to implementation limitations for existing tools, we used both the YOLO V3 \ac{CNN}~\cite{yolov3}, and two smaller size benchmarks, LeNet-5~\cite{lecun1989backpropagation} and CIFAR10~\cite{CIFAR10I1:online}.

With respect to the baselines, we compared accuracy and speedup of the proposed \ac{FES} solution against the \ac{FI} process carried out with SASSIFI; \ac{ourtool} achieves about 99\% accuracy in terms of effect of the fault on the final YOLO V3 output with about a 6x speedup. A second evaluation refers to the quality of the \ac{FES} approach with respect to the facilities offered by TensorFI, by comparing the available error models and the obtained speedup in executing the simulation on the same workload with the smaller \acp{CNN}, achieving a speedup from 44x to 63x.
While the speedup is a secondary aspect considering the rapid evolution and developments in \ac{FI} tools, it is the ease of use of the error models, the functional classification of the outputs and the final reliability analysis the elements we focus on.
We adopted \acp{GPU} as a platform, because they are the most commonly selected devices for accelerating \acp{CNN}; nonetheless we claim that the tools can be re-targeted for different devices, while the methodology remains the same.

The remainder of the paper is organized as follows. Section~\ref{sec:background} presents background on \acp{CNN} and on \acp{GPU}, while Section~\ref{sec:related} reviews the related work on \ac{FI} and \ac{FES} for \acp{CNN}. Section~\ref{sec:framework} discusses \ac{ourtool} in its details and Section~\ref{sec:modeling} presents the identified error models. Section~\ref{sec:simulation} then discusses the results of the application of our framework to a set of real-world \acp{CNN}, and the comparison against SASSIFI and TensorFI. Section~\ref{sec:adv} summarizes the main advantages of the methodology, and finally, Section~\ref{sec:lausdeo} concludes the paper.

\section{Background}
\label{sec:background}
This section provides a brief background introduction of the two key elements of the proposed application context, namely \ac{CNN} and \ac{GPU}.

\subsection{CNN}

\begin{figure}[t]
    \centering
    \includegraphics[width=\columnwidth]{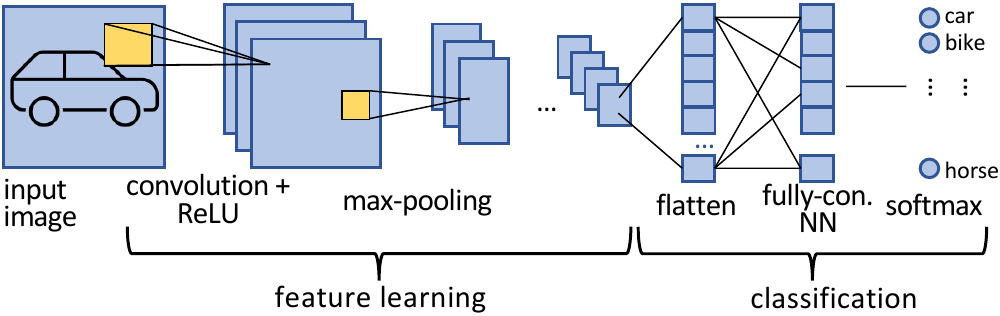}
    \caption{The typical topology of a \ac{CNN} described in terms of a sequence of layers for features learning and a final classification layer, integrating a fully-connected Neural Network.}
    \label{figure:cnn}
\end{figure}

A \acl{CNN}~\cite{lecun1989backpropagation} is a Deep Learning model generally employed in image processing and computer vision to derive a semantic representation from the input images to accomplish a high-end task, such as item classification, object detection and image segmentation.
As shown in Figure~\ref{figure:cnn}, a \ac{CNN} is internally organized in a sequence of \textit{layers}, each one processing multidimensional data, known as \textit{tensors}, by means of \textit{operators}. 
A tensor consists of a multi-dimensional stack of bi-dimensional matrices of values, called feature maps, generating a multidimensional grid. As an example, an image can be seen as three stacked feature maps, each one for a color channel, thus producing a 3D tensor.
There are several operators, that can be grouped into the following classes:
\begin{itemize}
    \item \textit{Convolution}, used to ``learn'' and extract features from the input by inferring the appropriate weights;
    \item \textit{Batch normalization}, used to \textit{fix} the data distribution, speeding up the learning process;
    \item \textit{Activation function}, a mathematical function applied element-wise to mimic the biological activation of a neuron. Common examples are the Sigmoid function, the softmax one and the \ac{RELU} one;
    \item \textit{Max-pooling} (and similar dimensionality reduction operators), used to reduce the size of the tensor for increasing the degree of generalization;
    \item \textit{Element-wise operators} for classical math operations or single-element manipulation, such as addition, multiplication, exponent and bias addition. 
\end{itemize}
Since operators are general in the size of the input/output tensors, they are characterized by a set of hyper-parameters, specifying the actual size of the processed tensors.
Operators belonging to these classes are usually organized in a sequence of layers devoted to the feature learning; each operator takes in input and produces in output tensors of specific sizes based on the structure of the \ac{CNN} (as shown in the example in left-hand side of Figure~\ref{figure:cnn}). The result produced by the feature learning is a tensor as well, that is considered as final output of the \ac{CNN} when the goal is for instance image segmentation or object identification. When the application goal is \textit{classification}, the \ac{CNN} contains a second part where the final tensor is first flattened and then is fed into a fully-connected Neural Network and a softmax function to produce the set of probabilities representing the likelihood of the identified object to be classified according to a set of classes.  

Given the complexity of designing a \ac{CNN} from scratch, a number of \ac{ML} design frameworks, such as TensorFlow~\cite{tensorflow2015-whitepaper}, Caffe~\cite{jia2014caffe}, PyTorch~\cite{paszke2019pytorch} and Keras~\cite{keras}, has been developed. They provide general and extensible programming interfaces in high-level languages (e.g., Python and C++) to easily specify the structure of a \ac{CNN} in terms of a dataflow model by instantiating operators available in the framework repository to create a graph. Moreover, these frameworks provide tools to automatically train the model to set the values for the parameters within each instantiated operator. Finally, they feature automated back-end support to target several processing devices such as CPUs or \acp{GPU} to optimize performance.

\subsection{GPU}
\begin{figure}[t]
    \centering
    \includegraphics[width=\columnwidth]{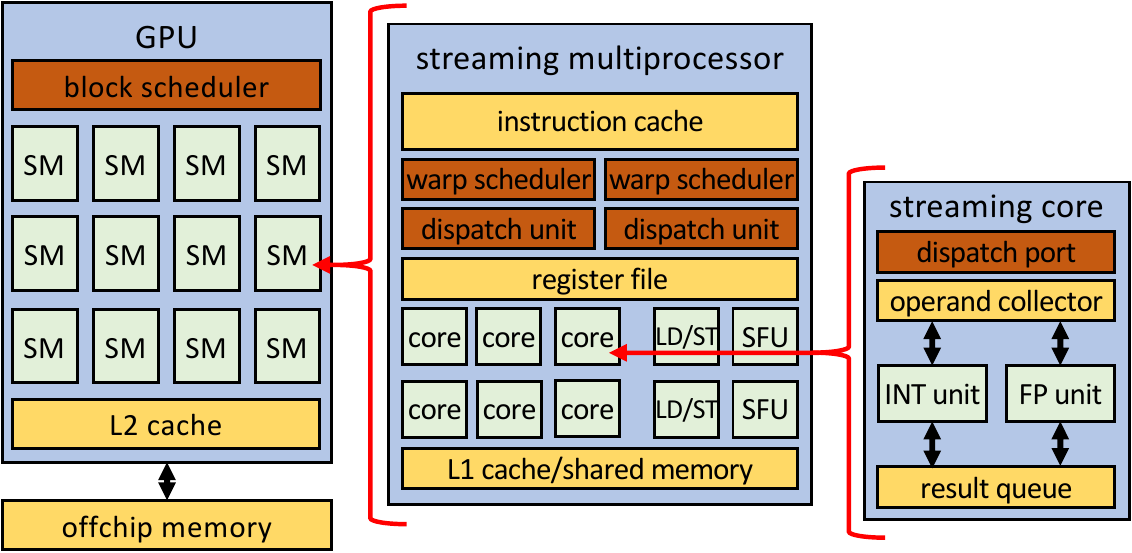}
    \caption{NVIDIA \ac{GPU} architecture hierarchically organized in a grid of streaming multicores; each multicore is a SIMD unit, having centralized scheduling and dispatching units and a set of processing elements such as streaming cores, load/store units (LD/STs) and special functional units (SFUs).}
    \label{figure:gpu}
\end{figure}

With reference to the NVIDIA architecture~\cite{lindholm2008}, a \ac{GPU} is a many-core organized in an array of \acp{SM} (left-hand side of Figure~\ref{figure:gpu}). The \ac{SM} is in turn structured following the \ac{SIMD} paradigm, thus with a single control unit scheduling and dispatching instructions, and a grid of simple arithmetic cores, memory load/store units and other special functional units for math operations, namely streaming cores, LD/STs and SFUs, respectively in the central part of Figure~\ref{figure:gpu}. Then, each single streaming core is internally implemented to read data from the register file (i.e. the operand collector), execute either integer or floating point operations and store results in the register file (result queue).

A \textit{kernel} is a software function accelerated on the \ac{GPU} by means of a large grid of threads grouped in a number of blocks. Threads compute on separate portions of data. Each threads' block is dispatched by the block scheduler to a single \ac{SM} for its execution; the group is subdivided into \textit{warps}, executed with the \ac{SIMD} paradigm; indeed the \ac{SM} contains multiple warp schedulers and dispatch units (2 in the figure) to execute concurrently several warps. Branch instructions, where divergences may occur among threads in the same warp, are executed by means of predicated approach; all alternative branches are executed in sequence for the entire warp and, each single thread is active only in the selected branch while it is disabled in the other ones.
Finally, the dispatching unit schedules several interleaving warps to maximize the \ac{SM} throughput.

The \ac{GPU} has a memory hierarchy organized in a off-chip global memory and two levels of caches, a unified L2 and a per-\ac{SM} L1. In addition to such a memory hierarchy, a \ac{SM} contains a shared memory, that is a local scratchpad memory that can be used internally by the threads of a single block to cooperate and synchronize without requiring longer delays.

\subsection{The fault model}
We consider the \acf{SEU} fault model, whose effect is a bit-flip in a value stored in a register of the computing platform.
Faults are rare events, therefore the classical assumption of a single fault per execution of the application holds. 
The occurrence of faults may cause the application:
\begin{inparaenum}[i)]
\item to crash or the operating system to raise an exception blocking the execution,
\item to hang leading to a non-termination, or
\item to terminate by producing an erroneous result, without any alert, a.k.a. \ac{SDC}~\cite{WT+14}.
\end{inparaenum}
While the first two cases are usually managed at operating system level, \acp{SDC} represent the most critical situation, and here we target this class of effects.

When considering the \ac{GPU} platform, as discussed in~\cite{sassifi}, an \ac{SEU} occurring in the \ac{SM} may corrupt the execution of a single thread causing an \ac{SDC}, or a group of threads within the same warp belonging to a single kernel; moreover, when shared memory is used, erroneous data produced by a thread may propagate among several threads within the same block of the kernel. On the other hand, memories and caches are hardened by means of \ac{ECC}; therefore, no fault in those locations will be here considered.

Within \ac{ML} design frameworks, each single operator is accelerated onto \ac{GPU} by means of one or multiple kernels. As a consequence, the final effect of a single \ac{SEU} affecting the \ac{GPU} running a \ac{CNN} is the corruption of an intermediate tensor, that forwarded to the subsequent layers will potentially produce an erroneous final result. Our framework aims at provide insights on such fault effects on \ac{CNN} execution.

\section{Related Work}
\label{sec:related}
A large effort has been devoted to the reliability analysis of \acp{CNN} as surveyed in~\cite{ibrahim2020}. To mention the most relevant works, several papers (e.g.~\cite{bosio2019,arechiga2018}) analyzed the effects of faults corrupting the weights of the convolutional layers on the final output to measure the overall resiliency of different \ac{CNN} models. \cite{reagen2018,LSS+2017} presented approaches tracing the error propagation through the layers to analyze application-level error masking and final effects on the final output, and to identify the most critical parts of the \ac{CNN}. In these works, faults corrupting both the memory storing the weights and the internal registers (causing effects on the final computation) have been considered. The conclusions are that different \ac{CNN} models exhibit different vulnerabilities depending on the adopted data types and number of layers.
In~\cite{santos2019} several metrics are defined to measure the vulnerability of the layers and kernels composing a \ac{CNN} model; then selective replication is applied to the most vulnerable parts to improve reliability while limiting overhead. A similar approach is proposed in~\cite{rech2019b} where an \acl{ABFT} technique is used to reduce the hardening costs. In~\cite{ibrahim2019}, selective \acl{TMR} is used; an interesting aspect of this solution is the idea of considering as critical only those faults that cause the overall application to perform a misclassification.
Finally, in~\cite{rech2016} an application-aware classification is performed by considering precision and recall on the obtained object detection outputs; selective duplication is also used to harden the most critical portions of the code.
In conclusion, due to the variety of possible internal structures and parameters of the \acp{CNN} and devices to accelerate such applications, the challenge is still open~\cite{ibrahim2019}.

As far as \ac{FI} facilities are concerned, radiation test offers the most adherent solution to cause soft errors, however, apart from the highly expensive equipment and the complex setup, there is little to no controllability on the injection process, and the observability is only on the primary outputs. As a result, it is suitable for final validation and black-box campaigns, not for in-depth analysis and investigations~\cite{rech2016,rech2019b}.  
To allow for controllability/observability in the campaigns, the most commonly adopted approach is emulating fault effects in the hardware platform running application, that is in the \ac{GPU} executing the \ac{CNN} in the present application scenario. GPU-Qin~\cite{fang2014gpu} exploits the CUDA-GDB debugger for NVIDIA devices to inject single bit-flips in the registers exposed by the \ac{ISA}; the solution is quite sophisticated and presents a 100x slowdown w.r.t. nominal execution. A similar debugging-based approach is used by CAROL-FI~\cite{rech2019}, which acts at the source code-level to inject both bit-flips and random values. The introduced performance degradation is below 5x, having both fault injection and error propagation analysis at source code-level. The limitation consists in the limited fault location sites, the used variables, introducing a significant gap between the \textit{injected errors} and the reality of soft errors affecting the hardware platform. SASSIFI~\cite{sassifi} and LLFI-GPU~\cite{li2016understanding} adopt a different strategy, by instrumenting the source code to support error injection, before executing it on the \ac{GPU}. SASSIFI, proposed by NVIDIA, can corrupt all \ac{ISA} registers through several injection modes, with a reported 5x slowdown~\cite{rech2019}. LLFI-GPU uses a similar approach, acting at the source code abstraction level, claiming a speedup of about 42x w.r.t. GPU-Qin, with similar benefits as CAROL-FI in terms of the analysis of the propagation of the errors. 

All these tools rely on complex modification and recompilation mechanisms to enable error emulation thus leading to a considerable performance degradation.
%but also, to the impossibility to be used. 
Some of these tools, such as GPU-Qin~\cite{fang2014gpu}, require a long profiling activity; moreover, as commented in~\cite{rech2019b}, SASSIFI is the only tool working with the NVIDIA proprietary libraries generally employed when implementing \acp{CNN}. Indeed, library requirements of the employed \ac{FI} tool, on the one hand, and of the considered application, on the other hand, may often conflict. As an example, \acp{CNN} implemented with the TensorFlow framework cannot be compiled for SASSIFI because they require different CUDA versions. Finally, the code instrumentation mechanism prevents the execution of complex applications; in our tests, a single run of YOLO V3 implemented in the Caffe framework and run in SASSIFI required more than 15 minutes, due to the large amount of data to be transmitted to the \ac{GPU}. Based on these considerations faster and more accurate tools are required, as it has also been stated in~\cite{ibrahim2020}.
The work in~\cite{NVBITFI}, contemporary to our work, proposes a new tool for NVIDIA \acp{GPU} called NVBitFI. The tool overcomes all limitations of previous solutions by performing a dynamic and selective code instrumentation to enable fault injection. NVBitFI does not require access to the source code, therefore improves both performance and usability. On the other hand, implementation issues still persist when working with external libraries.

Working at a higher abstraction level would be beneficial for two reasons: to dominate the complexity of \ac{CNN} applications, and to accelerate and facilitate experiments set-up and execution. To this end, simulation approaches have been proposed, injecting \textit{errors} directly in the execution state of the running \ac{CNN} application. 
The most representative example is TensorFI~\cite{chen2020}; it is an error simulator specifically tailored for \acp{CNN} integrated in the TensorFlow framework. The tool acts at the abstraction level of the application dataflow and injects errors by means of saboteurs that manipulate the output tensor. The tool so emulates the effects of faults affecting the \ac{CNN} operators execution. A similar strategy developed for other frameworks is the one presented in~\cite{reagen2018,MA+2020,neggaz2020}, integrated in the Keras, Pytorch and Caffe frameworks, respectively. \cite{arechiga2018,bosio2019} introduce two tools developed in Keras and Darknet, supporting only the corruption of the operator weights in the \ac{CNN}. Finally, the work in~\cite{ruospo2020} is implemented in PyTorch and is based on a cross-layer approach executing the \ac{CNN} model in software and switching down to the \ac{HDL} description of the processing unit for the actual injection. This very last approach has not been applied to \acp{GPU} due to the high complexity of the corresponding \ac{HDL} model, when available.

\begin{figure}[t]
    \centering
   \includegraphics[width=\columnwidth]{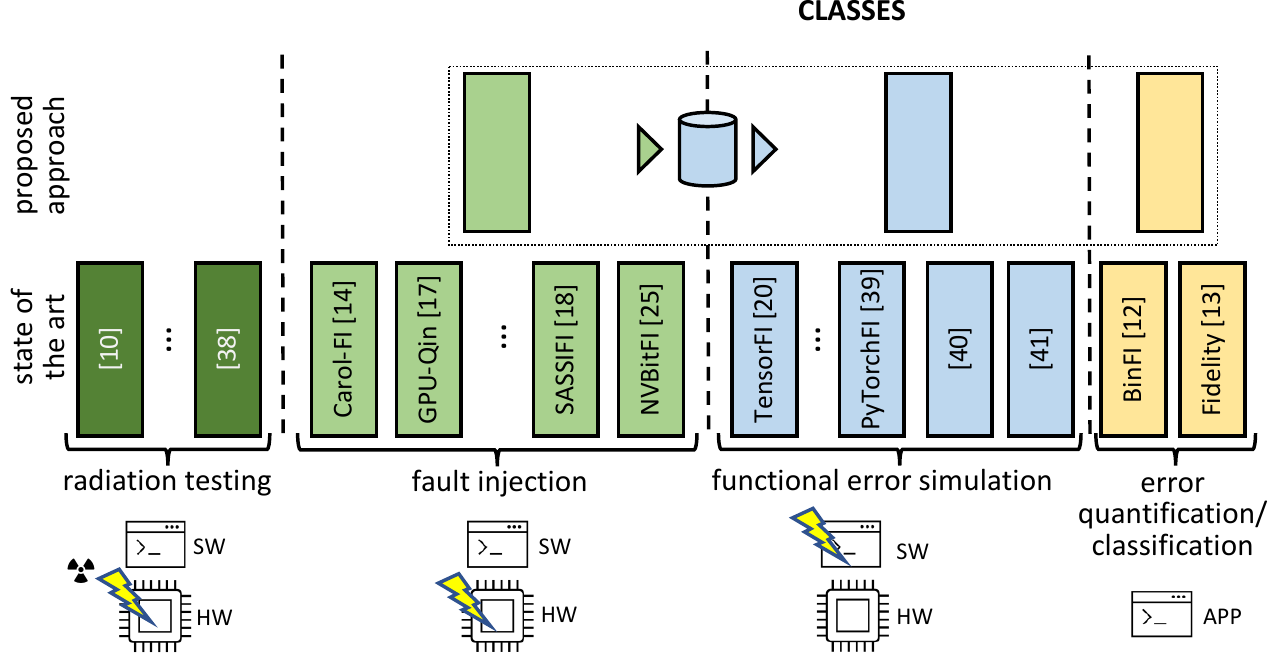}
 \caption{Comparison of the proposed approach w.r.t. the literature, highlighting the various contributions in radiation testing, architectural fault injection, functional error simulation, and error quantification/classification.}
    \label{fig:literature}
\end{figure}

The considerable advantage of \ac{FES} is that it can be easily integrated within the  \ac{ML} frameworks used in the design phase. Therefore, the \ac{CNN} reliability analysis and hardening can be performed from the early phases and in the same environment of its design and training. 
Moreover, this high-level approach decouples the reliability analysis from the actual execution on the final device (CPU or \ac{GPU} or custom hardware accelerator), thus sensibly easing the activity.
On the other hand, the main criticality when working at such high abstraction level, is the need of sound and complete error models, able to reproduce all and only the effects of physical faults occurring in the underlying hardware.
In~\cite{arechiga2018,bosio2019} only bit-flips in the weights are considered, neglecting the effects of faults affecting the processing unit, while in~\cite{reagen2018,neggaz2020} single or multiple bit-flips on the outputs of the \ac{CNN} operators are also considered. 
PyTorchFI~\cite{MA+2020} allows to inject random values, single bit-flips or zero values both in the weights and in the operator outputs; indeed, the paper explicitly states the necessity of defining more advanced error models describing the effects of faults affecting the underlying hardware. Finally, TensorFI adopts more advanced multiple-value corruptions, which sometimes poses problems to correlate the corruption effect to a real physical fault able to induce such an error, as we will discuss later in the paper.
Since the goal of the reliability analysis is to explore how a well designed and trained \ac{CNN} behaves when an unexpected problem in the underlying hardware occurs, it is fundamental to use error models that correspond to the effects caused by faults. In particular two risks are associated with error simulation: considering errors that cannot be the result of a real fault or ignoring effects of statistically relevant faults. As a consequence, the overall analysis might lead to inaccurate conclusions and tailoring hardening solutions that do not fit real needs.
Therefore, validated error models are vital to enable the adoption of error simulation.

Fidelity~\cite{HBL2020} and BinFi~\cite{BINFI} are the most recent works similar to our proposal. The former defines suitable error models adherent to what is observed through \ac{FI} and analyses the effects of the faults in terms of the \ac{AVF}. The latter focuses on identifying the safety-critical bits in ML applications, relying on TensorFI for the injection of bit-flips in the operators. The effects of the faults on the outputs are analyzed with respect to the application, to determine the actual impact in the case of safety-critical contexts. Our framework integrates and merges both strategies, extracting different but comparable error models w.r.t. Fidelity, and performing a classification of the effects similar to the one in BinFI, yet based on a different error injection solution with the limitations of the existing \ac{FES} based on TensorFI. In doing so, our proposal fills the gap between \ac{FI}, \ac{FES} and an application-related resiliency classification. As future work, we will investigate how Fidelity's error models can be used in CLASSES, and how CLASSES's final output can be re-formulated in terms of the metric adopted by BinFI.

Figure~\ref{fig:literature} shows a graphical summary of the reviewed literature and positions our cross-layer proposed approach. It exploits \ac{FES}, integrating the facility in the most popular \ac{ML} framework and adopting sound error models opportunely extracted by means of accurate low-level \ac{FI} campaigns on the target hardware platform. 

\section{Methodological Framework}\label{sec:framework}
\begin{figure}[t]
    \centering
    \includegraphics[width=\columnwidth]{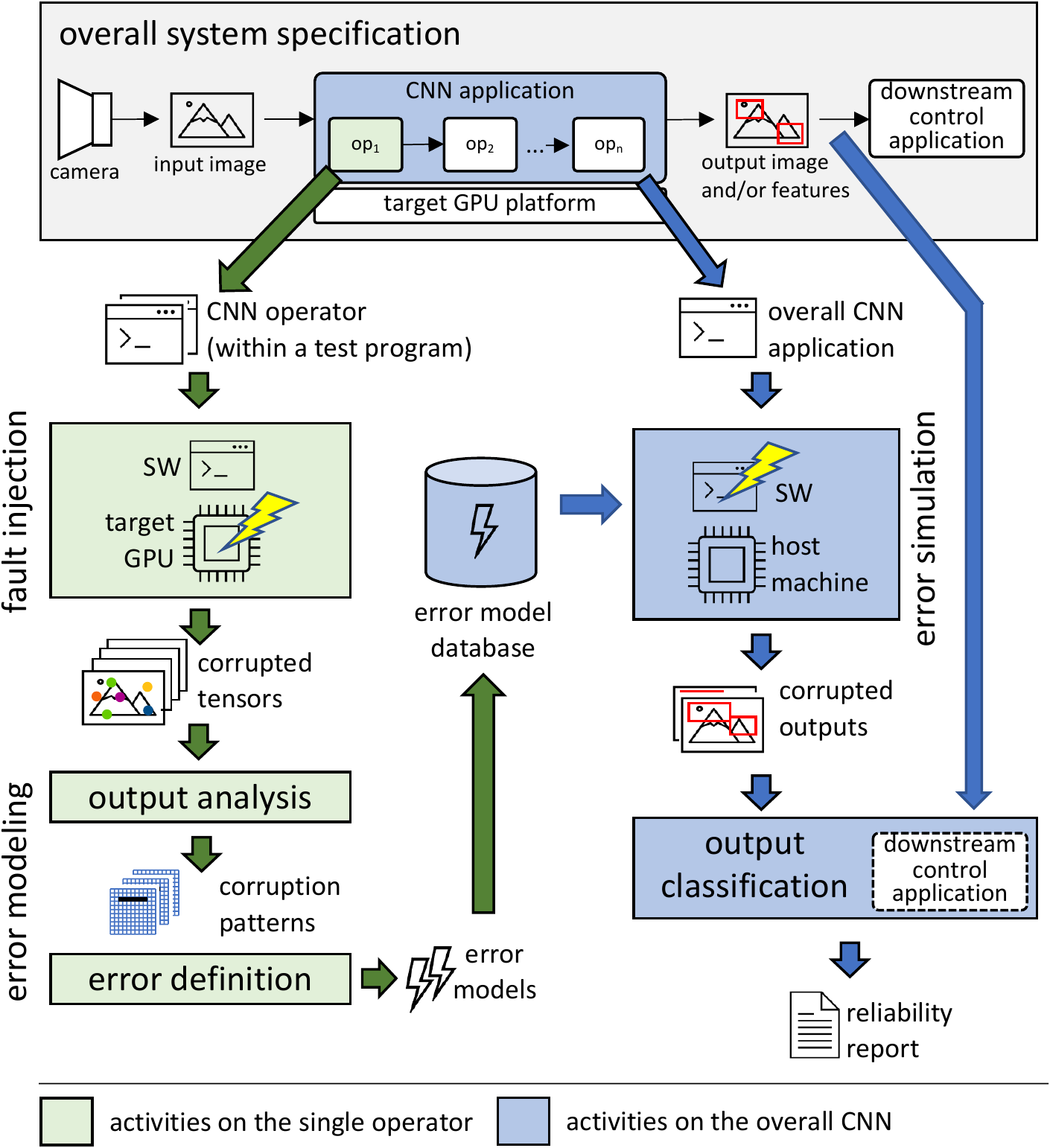}
    %\caption{\ins{The proposed cross-layer reliability analysis framework. The top-most part of the figure describes the overall system composed of the camera, the \ac{CNN} under analysis and the downstream control application. The subsequent part of the figure can be split horizontally. The left part presents the fault injection and error modeling flow, applied on each CNN operator to define the related error models. The right part presents the error simulation that is applied on the entire \ac{CNN} application; the subsequent output classification is used to perform an usability-based analysis of the results of each run of the application.}}
    \caption{The proposed cross-layer reliability analysis framework: the system under analysis (top most part), the fault injection and error modeling part applied at the granularity of the single CNN operator (green left side) and the error simulation environment for the overall usability-based reliability analysis (blue right side).}
    \label{fig:flow}
\end{figure}

A high level representation of the proposed cross-layer reliability analysis framework, \ac{ourtool}, is depicted in Figure~\ref{fig:flow}. The framework has been designed for \acp{CNN}, and more in general for Deep Learning applications, accelerated onto a target \ac{GPU}-based platform. \acp{CNN} are generally exploited for perception tasks; therefore, we here consider a larger system where the \ac{CNN} takes images from a source, such as a camera, and its output, being an enhanced image or a set of features, is transmitted to a downstream application using them for some decision making task.

The input of the framework is the \ac{CNN} application under analysis implemented as a graph of operators op$_1$, op$_1$, ..., op$_\mathtt{n}$ in the adopted \ac{ML} design framework, e.g., TensorFlow~\cite{tensorflow2015-whitepaper}, and already properly trained, and the output is a detailed report of the performed reliability analysis highlighting the vulnerabilities and the weaknesses of the application under design.
The framework has a cross-layer structure since it mixes architecture-level \ac{FI} and application-level \ac{FES} with to take advantage of the benefits of both. In particular, \ac{FI} offers high accuracy in emulating faults in the target \ac{GPU} hardware while \ac{FES} is flexible and fast in simulating fault effects directly in the application without requiring the specific target platform to be deployed and instrumented.
To this end the framework is divided into two main parts as shown in Figure~\ref{fig:flow}, discussed in details in the following subsections.

\subsection{Fault Injection and Error Modeling}\label{sec:method_errormodeling}
Fault injection is here used to analyze the output of the \ac{CNN} basic operators when their execution on the target platform is corrupted by the occurrence of a fault in the underlying hardware. The final goal of this activity is to define a set of error models reproducing the effects of hardware faults on the output tensor of the executed operator.

The list of all operators used in the considered \ac{CNN} is extracted and each operator is individually analyzed.
The stand-alone operator is wrapped within a test program transmitting the input tensor and collecting the output one. For the sake of accuracy of the analysis, the values in the input tensor are defined according to the intermediate tensors extracted from a run of the overall \ac{CNN} application. Moreover, several input tensors are considered to avoid any data-dependent bias in the obtained result.
The test program is used for a large architecture-level fault injection campaign aimed at collecting a large set of corrupted output tensors.

The obtained corrupted tensors are then analyzed to derive a set of accurate high-level \textit{functional error models} representative of all possible effects of hardware faults in the specific operator. 
In details, since a tensor is a multidimensional matrix of numeric values, each corrupted tensor is compared against the golden counterpart according to three different aspects:
\begin{itemize}
\item \textit{error cardinality}, i.e. the number of  erroneous values in the tensor, 
\item \textit{error value domain}, i.e. the domain the erroneous values belong to, and
\item \textit{error spatial distribution}, i.e. how erroneous values are distributed in the multidimensional matrix.
\end{itemize}
It is worth mentioning that these three aspects, and particularly the last one, are highly related to the peculiar characteristics of the \ac{GPU} architecture based on the \ac{SIMD} paradigm.
Based on these aspects, the corrupted tensors are inspected to identify recurrent \textit{corruption pattern}s and their occurrence frequency.
The identified corruption patterns are also studied to assess if they are independent of the specific input sample and if they can be ``reproduced'' by an algorithm applied to the golden tensor. If all these conditions hold, the corruption pattern leads to the definition of an error model, that is implemented in the framework as a saboteur executed on the output of the corresponding operator.
This activity, performed on all the operators, allows to build a database of error models to be then exploited to perform the reliability analysis of the overall \ac{CNN} by means of an application-level error simulation.

\ac{ourtool} presents several advantages.
As mentioned, the adoption of architecture level fault injection offers a high accuracy in the obtained results. At the same time, we are analyzing the stand-alone operators instead of the entire application simplifying the overall task.
We operate on the single operators, that are the basic blocks of the \ac{CNN} design, analyzing the input/output relation due to the presence of the fault.
Moreover, many operators, such as the convolution, present hyper-parameters defining the size of the input and output tensors. For each of them, the hyper-parameters are tuned to execute fault injection and error modeling on the smallest version of the operator that allows for obtaining error models valid for any other version of the same operator.
As we will see later, these choices considerably reduce the set up effort and the execution time of the fault injection campaign without compromising its soundness and completeness.

Finally, since \acp{CNN} are generally composed of a recurrent set of the same operators, when a new application is used, only the operators not already analyzed in previous campaigns need be taken into account, as their fault-error relation is application independent.

\subsection{Error Simulation and Output Classification}
\label{sec:classification}
The overall \ac{CNN} application is analyzed w.r.t. fault effects by means of error simulation. According to the defined error models, the granularity of the corruption is at \ac{CNN} dataflow graph level, by considering operators as elementary operations and corrupting the output tensors.
The adopted error simulation strategy is thus based on saboteurs introduced between two nodes of the \ac{CNN} dataflow graph corrupting the output tensor of the source operator.
Thanks to the flexibility of the \ac{ML} framework, \ac{CNN} dataflow graph structural analysis and instrumentation can be automatically performed. Similarly, probes can be inserted to trace error propagation.
Finally, error simulation follows a pretty standard execution workflow: 
\begin{inparaenum}[i)]
\item \ac{CNN} structural analysis and instrumentation;
\item error list generation; 
\item error simulation for each item in the list, collecting and classifying the produced outputs; and
\item final reliability report generation.
\end{inparaenum}

The error simulation approach presents higher flexibility, usability and performance than the classical fault injection counterpart. From the implementation point of view the structure of the software is less complex with less implementation and compilation issues, thus also obtaining better performance.
Moreover, error simulation can be carried out on any host machine, non-necessarily featuring the target \ac{GPU}; in fact, the adopted error models already represent the effects of the faults on such a platform. Thus, large workstations can be adopted to reduce the execution times of the error injection campaigns. 
From a methodological point of view, the approach allows for a greater controllability and observability of the fault effects also because error simulation does not suffer from fault activation issues, further reducing execution times.

Another relevant aspect of the proposed framework is related to how error simulation results are analyzed. 
As discussed in~\cite{BBC+2020}, the classical strategy based on a bit-wise comparison of the results against the golden counterpart to classify the experiment as \textit{correct vs. error} is here not effective due to the approximate and inexact nature of \ac{CNN} applications. 
Therefore, we adopted a \textit{usability-based classification}; by considering the fact that the \ac{CNN} is part of a larger system (refer to Figure~\ref{fig:flow}). The \textit{output classification} module is in charge of determining whether the corrupted output produced by the \ac{CNN} still allows the downstream control application to perform its elaborations in an acceptable way or not. This is possible because the output classification module integrates the logic of the downstream control application itself and an application-specific policy that is actually meant to asses the usability of the produced output. As an example, we may consider a CNN performing an object detection task that supports an autonomous object grabbing robot: a slightly shifted bounding box in the produced output may be considered admissible since the robot may still be able to grab the object; therefore, in this case the corrupted output would be classified as \textit{usable}. Conversely, a missing bounding box would cause the robot not to grab the object, and thus this second corrupted output would be classified as \textit{unusable}.
The advantage of this approach is therefore to focus on faults having a disruptive effect, identifying the main vulnerabilities of the \ac{CNN} and its critical components, ignoring the ones that have a limited impact, inherently tolerated by the nature of the \ac{CNN} computation.

\section{Framework Implementation}\label{sec:implementation}
\begin{figure}[t]
    \centering
    \includegraphics[width=.9\columnwidth]{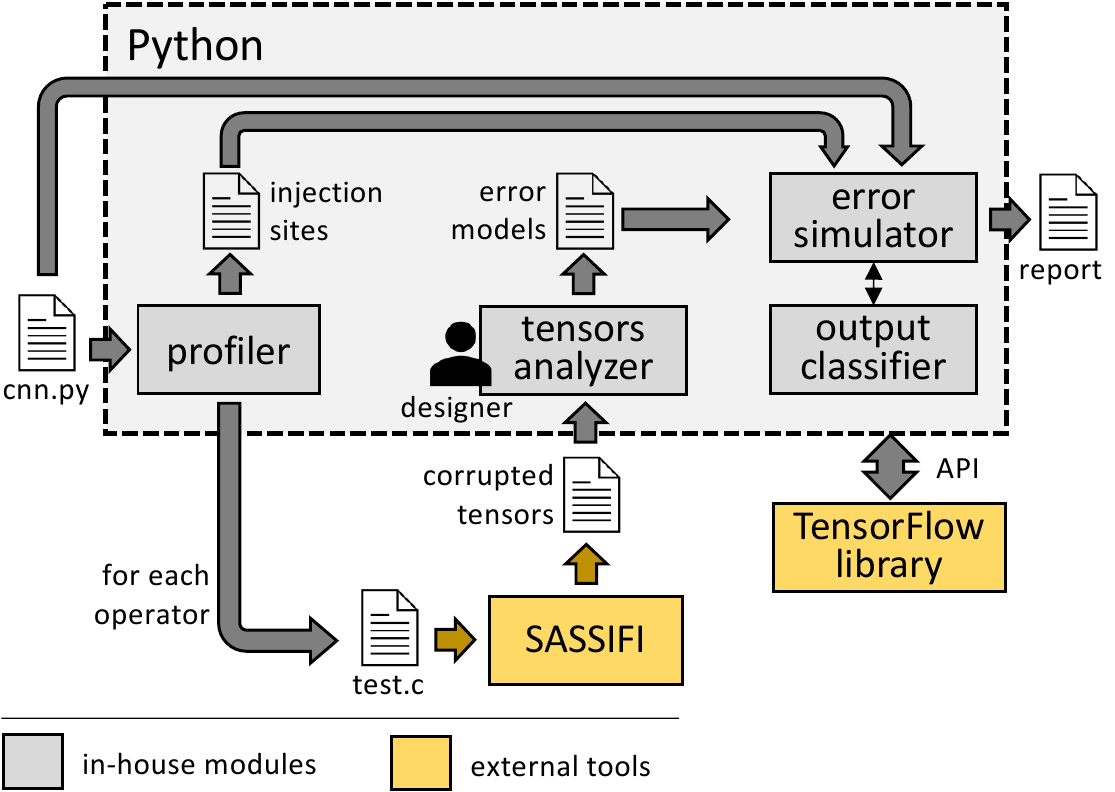}
    \caption{\ac{ourtool} internal organization.}
    \label{fig:tool}
\end{figure}

\ac{ourtool} has been implemented as a semi-automated tool in Python and integrated within the TensorFlow framework, accessing the SASSIFI fault injector as an external module. The tool is semi-automated in the sense that almost all steps are automatically executed; the role of the designer is to supervise the overall workflow and critically check the correctness and the quality of the outputs of each step.
The resulting structure is shown in Figure~\ref{fig:tool}, where the \textit{profiler} analyzes the considered \ac{CNN} and identifies the fault injection sites for each operator. The \textit{tensor analyzer} isolates the erroneous output tensors coming from the fault injection experiments and, supervised by the designer, it extracts recurrent corruption patterns within the analyzed tensors. The \textit{error simulator} performs the functional error simulation of the whole \ac{CNN} and, finally, the \textit{output classifier} determines whether the outputs of the considered \ac{CNN} after error simulation would be usable by the downstream application or not. Internals are discussed in the following subsections.

\subsection{Fault Injection and Error Modeling}
A first step (carried out by the \textit{profiler} module in the software structure shown in Figure~\ref{fig:tool}) is a pre-processing automatically performed by a set of Python functions using the TensorFlow \ac{API}.  
This step instantiates an instrumented copy of the graph used to perform a preliminary run of the \ac{CNN}. This activity allows for extracting the list of the employed operators, their hyper-parameters and some input/output tensors tuples from the \ac{CNN} graph.
The identified list of operators used in the \ac{CNN} under analysis is filtered to keep only  those operators for which error models are not available yet. If an operator is used in various versions, the one with smallest hyper-parameter values is considered.

The fault injection activity uses SASSIFI. The choice of keeping this part separate from the rest of the framework is that \ac{FI} is the only activity that has to be executed on the specific platform targeted by the application scenario. At the opposite, the main part of the framework is platform-independent, and it can be executed on any machine. 

Test programs for the identified operators are implemented in C++ by using Caffe~\cite{jia2014caffe} since TensorFlow cannot be integrated in SASSIFI due to library conflicts. However, there is a 1:1 correspondence between SASSIFI and Caffe operators thus not leading to any methodological issue in our approach.
Nonetheless, this implementation choice considerably simplifies the development and integration in SASSIFI; Caffe is a stand-alone library accelerated onto \ac{GPU}.
The test program of each operator is automatically generated by the tool by using a source code template. Basically, the test program loads the input tensor, executes the operator onto the \ac{GPU}, and saves the output tensor. For each operator, the tool specializes the template with the hyper-parameters, weights and input tensor extracted by the profiler. Finally, SASSIFI is configured to inject single bit-flips (to represent \acp{SEU}) in all available sites and the execution is completely automated.

The output of the fault injection campaign executed on each operator is processed within the Python framework to perform the analysis based to the three aspects discussed in Section~\ref{sec:method_errormodeling}, thus classifying results in clusters (this activity is carried out by the \textit{tensors analyzer} module in the structure shown in Figure~\ref{fig:tool}). The most challenging aspect is the analysis of the error spatial distribution that would require the designer to manually inspect the produced output. However, based on the peculiarities of the \ac{GPU} architecture and of the operators code, errors are mainly distributed in regular patterns (e.g., lines in the same feature map or lines crossing feature maps). Therefore, a manual inspection is here only required to classify corrupted tensors not handled by the automated script. 
It is worth noting that this manual inspection is the only step which requires the designer interaction (as shown in the figure); future work will be devoted in its automation by means of data mining techniques.

The output of this step is a tabular report describing all the identified clusters together with the corresponding occurrence frequencies.
This information is saved in a JSON file constituting the error model database.

\subsection{Error Simulation and Output Classification}

\begin{figure}[t]
    \centering
    \includegraphics[width=\columnwidth]{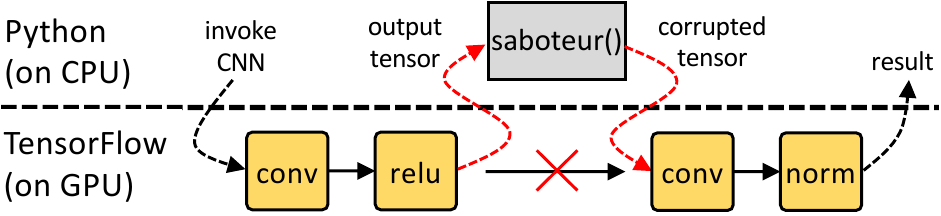}
    \caption{Error injection strategy. The sequence of operators in the \ac{CNN} are accelerated onto the \ac{GPU} until the operator to be corrupted is reached. The output tensor of this operator is fed into a saboteur (executed on a host CPU) that corrupts it by applying an error model. The corrupted tensor is then used as input of the remaining part of the operators (again, accelerated onto the \ac{GPU}) to complete the execution of the \ac{CNN}.}
    \label{fig:saboteur}
\end{figure}

\ac{ourtool} exploits the TensorFlow public \ac{API} to load and execute the \ac{CNN} model and to perform injection activities. It uses the information extracted during the pre-profiling activity to define the list of all injection sites; in particular, each output tensor is considered as a candidate.

The error injection strategy, carried out by the \textit{error injection} module in the software structure shown in Figure~\ref{fig:tool}, is implemented by virtually splitting the graph into two parts based on the injection site. As shown in Figure~\ref{fig:saboteur}, the first part of the graph is executed to compute the tensor where the error has to be injected, the saboteur is executed and the corrupted tensor is re-introduced in the second part of the graph to compute the final \ac{CNN} output. In other words, the insertion of a saboteur that modifies a tensor at the output of an operator allows to emulate the effect of the occurrence of a fault in the hardware platform when executing that operator while computing the modified tensor.

The saboteur is a Python function executed between two \ac{CNN} operators, featuring the error models. As discussed in Section~\ref{sec:modeling}, many operators share a similar algorithmic implementation; therefore, several corruption patters are common among different operators, while their occurrence frequencies may vary. Therefore, the saboteur has been designed to implement all the identified corruption patters and the JSON file provides the fundamental information mapping patterns to operators and occurrence frequency.

The final step of the proposed error simulation environment is carried out by the \textit{output classifier} module in the software structure shown in Figure~\ref{fig:tool}. As introduced in Section~\ref{sec:classification}, this activity consists in determining whether the produced corrupted output will be usable of not by the end-user application according to the working scenario.

The error simulator implements a standard workflow: 
\begin{inparaenum}[i)]
\item error list generation according to the injection sites identified by the \ac{CNN} pre-processing and designer parameters (e.g., the number of experiments);
\item experiment execution, i.e. run a number of error injection experiments, each one injecting in a randomly selected operator a randomly selected error while executing the application;
\item output classification by means of a user-defined classifier (a Python script) that assesses the usability of the output, according to the working scenario.
\end{inparaenum}
Typically the number of injection sites is quite limited, therefore a caching strategy has been actuated to optimize the execution times of the tool; in particular, the tensor to be corrupted is cached so that for all experiments on the same injection site, the first part of the graph is not re-executed.

As a final note, the overall prototype consists of approximately 2,000 lines of codes.

\section{Error Modeling Results}
\label{sec:modeling}
This section presents the results of the application of the methodological framework in the definition of the error models.

\subsection{Experimental Setup}
\label{sec:setup1}
As a real-world case study for our error modeling activity we considered 
the YOLO V3 \ac{CNN}, the state-of-the-art solution for object detection, also employed in commercial autonomous driving systems such as Apollo~\cite{Apollo17:online} and Autoware~\cite{Autoware68:online}.
The considered implementation was trained upon the COCO dataset~\cite{lin2014microsoft}. 
YOLO V3 internal structure counts about 6,000 operators instances belonging to 45 different operator types. The target \ac{GPU} was the NVIDIA GeForce GT 750M implementing the Kepler architecture.

\begin{figure*}[t]
    \centering
    \includegraphics[width=\textwidth]{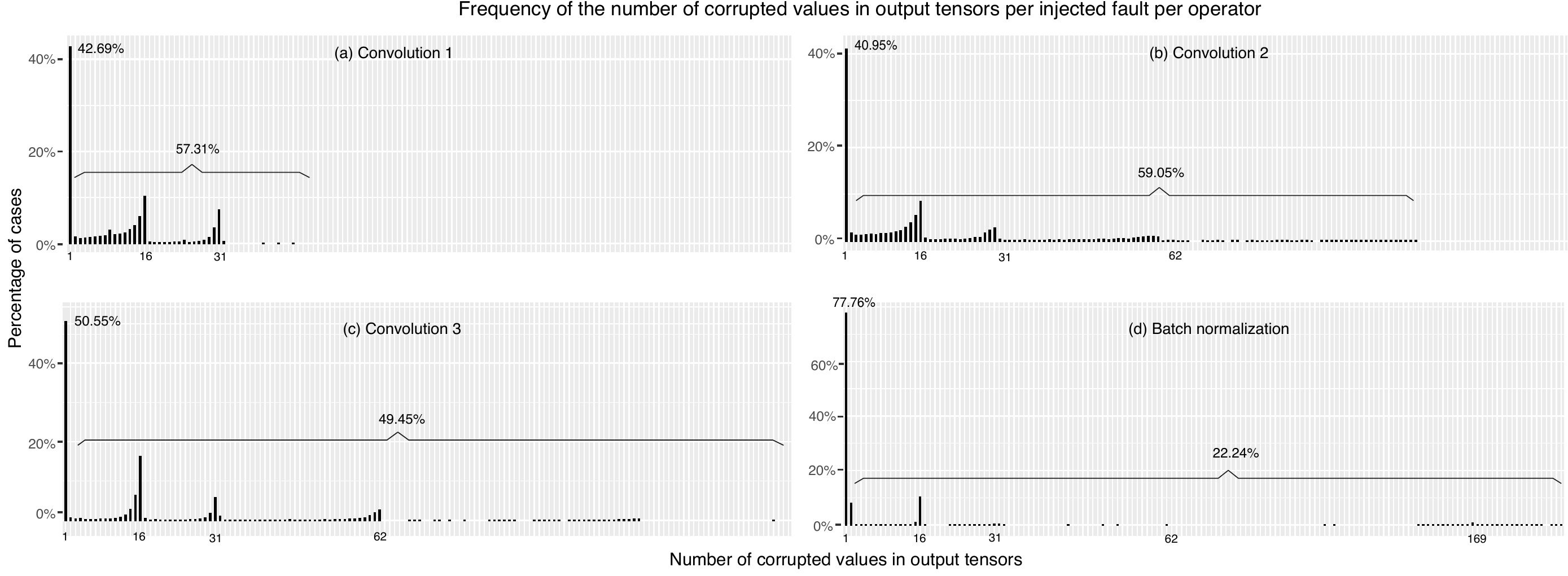}
    \caption{Frequency of the number of erroneous values per output tensor for different operators, highlighting the overall $\%$ of cases where multiple values are corrupted (e.g., 57.31$\%$ for \texttt{Convolution 1}).}
    \label{figure:cardinality}
\end{figure*}

\begin{table}[tb]
    \caption{YOLO V3 operators considered for \ac{FI} \label{table:operators}
}
    \centering
\resizebox{\columnwidth}{!}{%
    \begin{tabular}{|l|c|c|c|}
        \hline 
        \multirow{2}{*}{\textbf{Operator}} & \multirow{2}{*}{\textbf{Input size}} & \multirow{2}{*}{\textbf{Output size}}& \textbf{Additional} \\
         &  &  & \textbf{parameters} \\        \hline
        \texttt{Conv. 1} & $512 \times 13 \times 13$ & $256 \times 13 \times 13$ & K = 1, Str = 1\\
        \hline
        \texttt{Conv. 2} & $128 \times 52 \times 52$ & $256 \times 52 \times 52$ & K = 3, Strt = 1\\
        \hline
        \texttt{Conv. 3} & $256 \times 52 \times 52$ & $512 \times 26 \times 26$ & K = 3, Str = 2\\
        \hline 
        \texttt{Add} & $1024 \times 13 \times 13$ & $1024 \times 13 \times 13$ & \\
        \hline
        \texttt{Batch Norm.} & $256 \times 13 \times 13$ & $256 \times 13 \times 13$ & \\
        \hline
        \texttt{Bias add} & $256 \times 13 \times 13$ & $256 \times 13 \times 13$ &\\
        \hline
        \texttt{Div} & $1 \times 10647$ & $1 \times 10647$ & \\
        \hline
        \texttt{Exp} & $1 \times 8112 \times 2$ & $1 \times 8112 \times 2$ & \\
        \hline
        \texttt{Leaky ReLU} & $256 \times 26 \times 26$ & $256 \times 26 \times 26$ & Slope = 0.1 \\
        \hline
        \texttt{Mul} & $1 \times 8112 \times 2$ & $1 \times 8112 \times 2$ & \\
        \hline
        \texttt{Sigmoid} & $1 \times 2028 \times 80$ & $1 \times 2028 \times 80$ & \\
        \hline
    \end{tabular}}
\end{table}

Table~\ref{table:operators} summarizes the operator instances considered in the \ac{FI} experiments and reports the operator name together with the size of the input and output tensors. For some operators additional parameters are reported, namely the adopted kernel size and stride for the convolutions and the negative slope value for the \texttt{Leaky ReLu}.

\subsection{Results from Fault Injection Experiments}
We selected the single bit-flip fault among the models supported by SASSIFI. Due to the complexity of modern \acp{GPU} and to the size of the their memory, an exhaustive \ac{FI} campaign is unfeasible. Nevertheless, it is possible to exploit the extreme regularity of the \ac{SIMD} architecture to reduce the number of required experiments. Indeed, several threads simultaneously executing the same code, elaborate on different bunch of data of the same type and are run on different instances of the same hardware resources. Thus, the output corruption patterns observed when injecting faults during the execution of one of the threads will be representative of the effects of faults in all threads. The size of the fault list for each operator has been determined to inject in all memory and register bits accessed by a single thread for each instruction of the execution trace of the operator when executed on the \ac{GPU}. When considering the high regularity of the \ac{SIMD} architecture of a \ac{GPU}, the setup allows one to obtain the same statistical results of an almost exhaustive injection campaign in the entire architecture. 
To make our \ac{FI} experiments as general as possible, in each experiment we changed the input of the operator and we randomly selected the fault-injected thread. The outputs of the \ac{FI} campaign are collected and compared against the expected output. Uncorrupted outputs are discarded while erroneous ones are further inspected to identify the recurrent error patterns. Overall we ran 360,000 \ac{FI} experiments and collected 137,004 corrupted tensors; the number of injected faults and collected corrupted tensors for each operator are reported in the second and third columns of Table~\ref{table:FI}, respectively.

\begin{table}[t]
    \caption{Results from the fault injection experiments    \label{table:FI}
}
    \centering
%\resizebox{.9\columnwidth}{!}{%
    \begin{tabular}{|l|r|r|}
        \hline 
        \multirow{2}{*}{\textbf{Operator}} & \textbf{Number of\phantom{00}} & \textbf{Number of\phantom{000}}  \\
         & \textbf{injected faults} & \textbf{corrupted tensors}  \\        \hline
        \texttt{Conv. 1} & 56,000 & 24,273 \\
        \hline
        \texttt{Conv. 2} & 38,000 & 15,488 \\
        \hline
        \texttt{Conv. 3} & 66,000 & 31,245 \\
        \hline 
        \texttt{Add} & 16,000 & 5,900 \\
        \hline
        \texttt{Batch Norm.} & 88,000 & 26,182 \\
        \hline
        \texttt{Bias add} & 16,000 & 7,400\\
        \hline
        \texttt{Div} & 16,000 & 4,400 \\
        \hline
        \texttt{Exp} & 16,000 & 6,400 \\
        \hline
        \texttt{Leaky ReLU} &  16,000 & 5,100 \\
        \hline
        \texttt{Mul} & 16,000 & 5,700 \\
        \hline
        \texttt{Sigmoid} & 16,000 & 4,500 \\
        \hline
    \end{tabular}
    %}
\end{table}

\subsection{Error Modeling}
We systematically analyzed the 137,004 collected corrupted tensors to identify the recurrent corruption patterns.
In the following we report the results of such analysis based on the previously discussed aspects of interest (error cardinality, error value domain and error spatial distribution), followed by the definition of the error models.

\begin{figure*}[t]
    \centering
    \includegraphics[width=.9\textwidth]{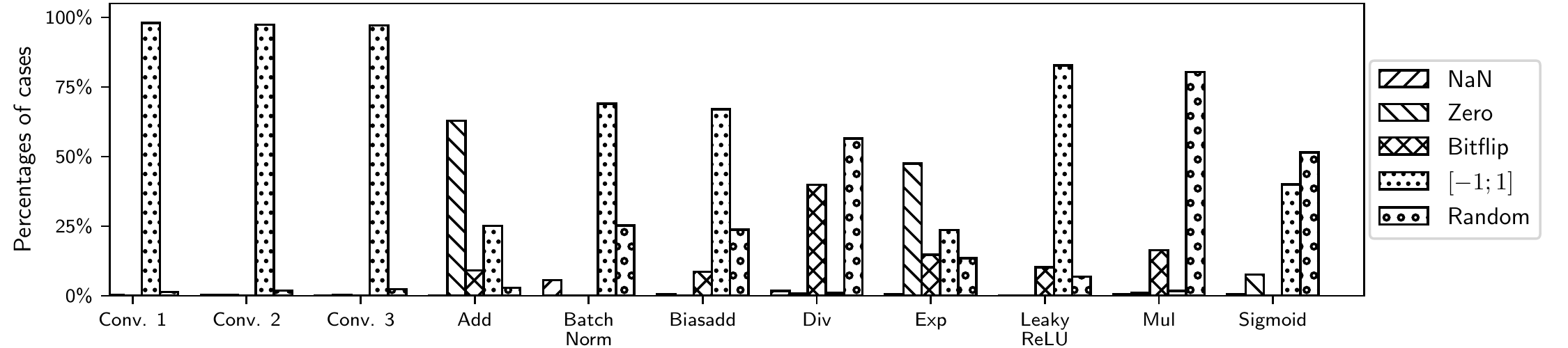}
%    \caption{Distribution of erroneous values domains in corrupted tensors per operator.}
    \caption{Distribution of erroneous values per each considered operator. For each operator, erroneous values in the corrupted output tensors have been classified as NaN, Zero, BitFlip, $[-1;1]$, or Random.}
    \label{figure:domains}
\end{figure*}

\begin{figure*}[t]
  \centering%
  \subfigure[Single point\label{figure:smf_sp}]%
    {\includegraphics[width=.26\columnwidth]{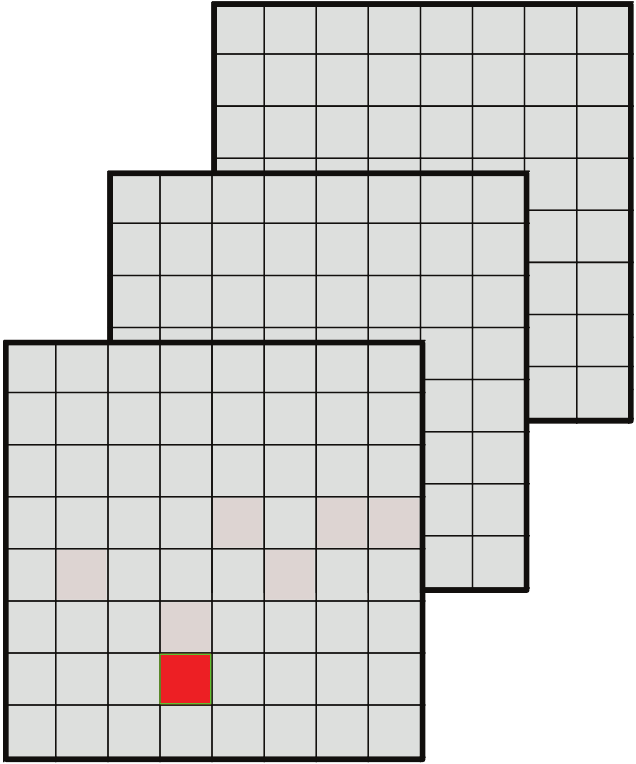}
  \includegraphics[width=.26\columnwidth]{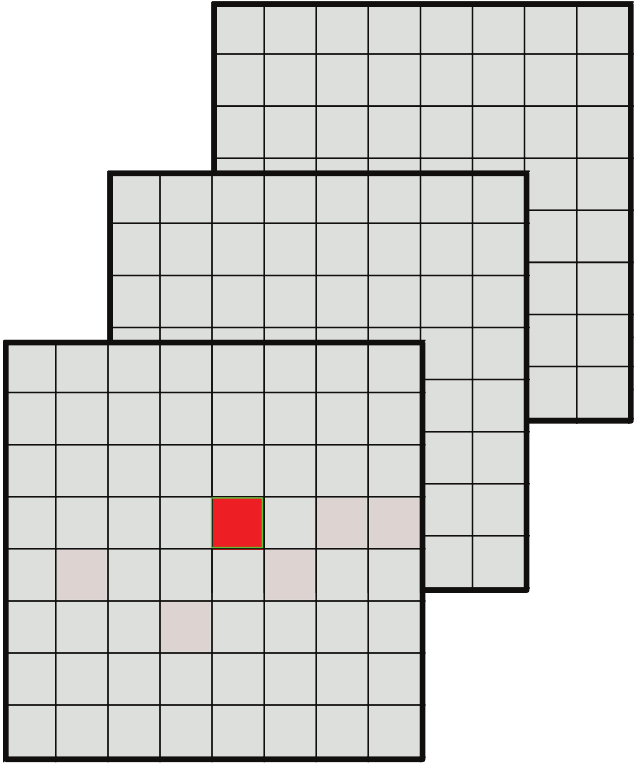}
  \includegraphics[width=.26\columnwidth]{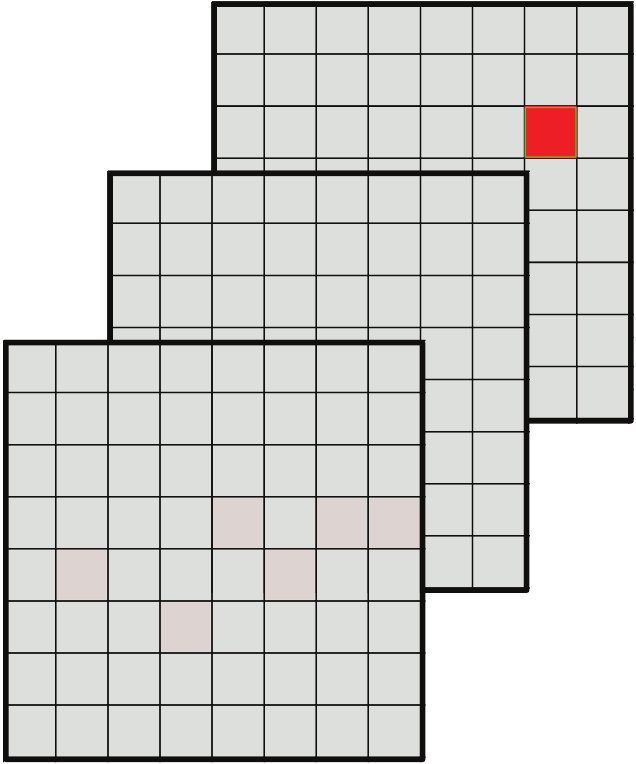}}
  \qquad
  \subfigure[Same row\label{figure:sameRow}]%
  {\includegraphics[width=.26\columnwidth]{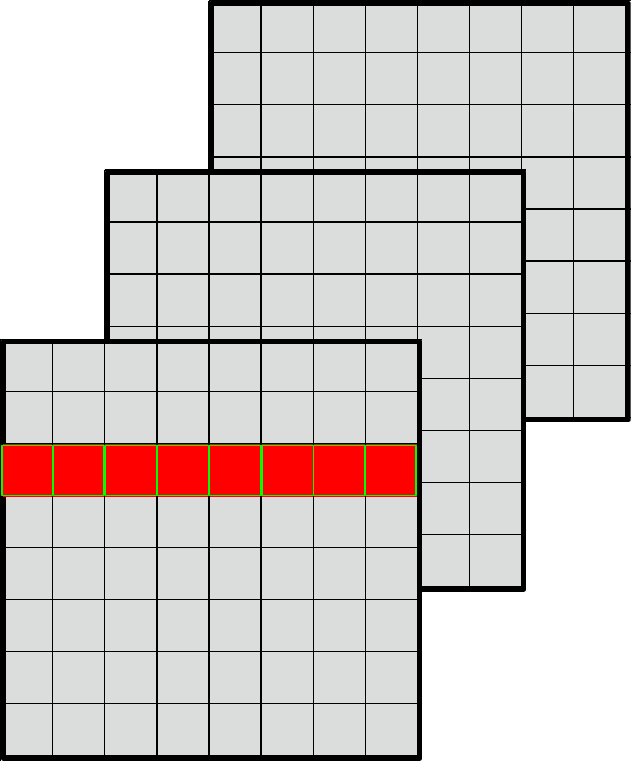}
  \includegraphics[width=.26\columnwidth]{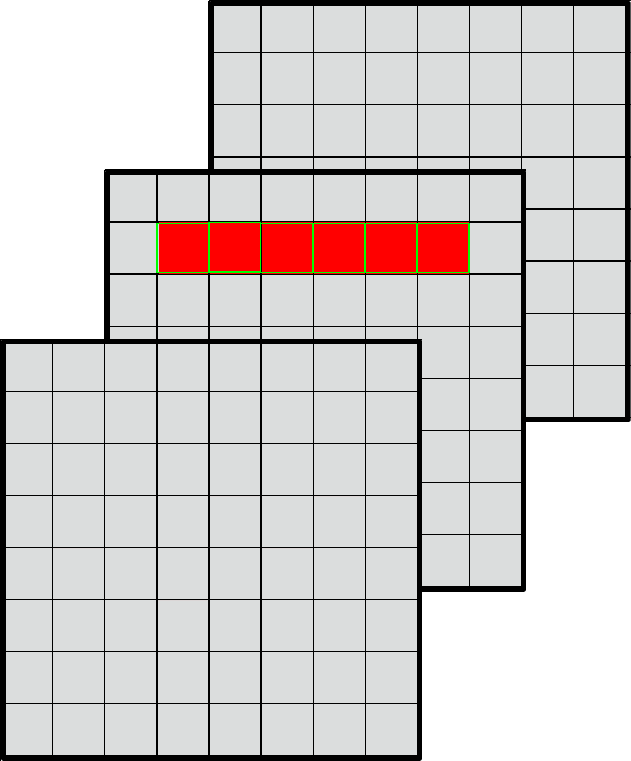}
  \includegraphics[width=.26\columnwidth]{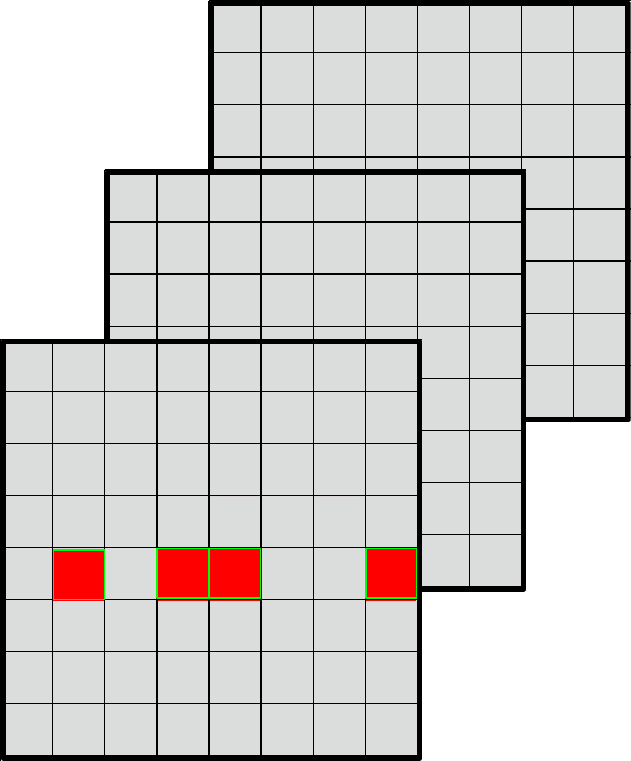}}
  \qquad
  \subfigure[Bullet wake\label{figure:bullet}]%
  {\includegraphics[width=.37\columnwidth]{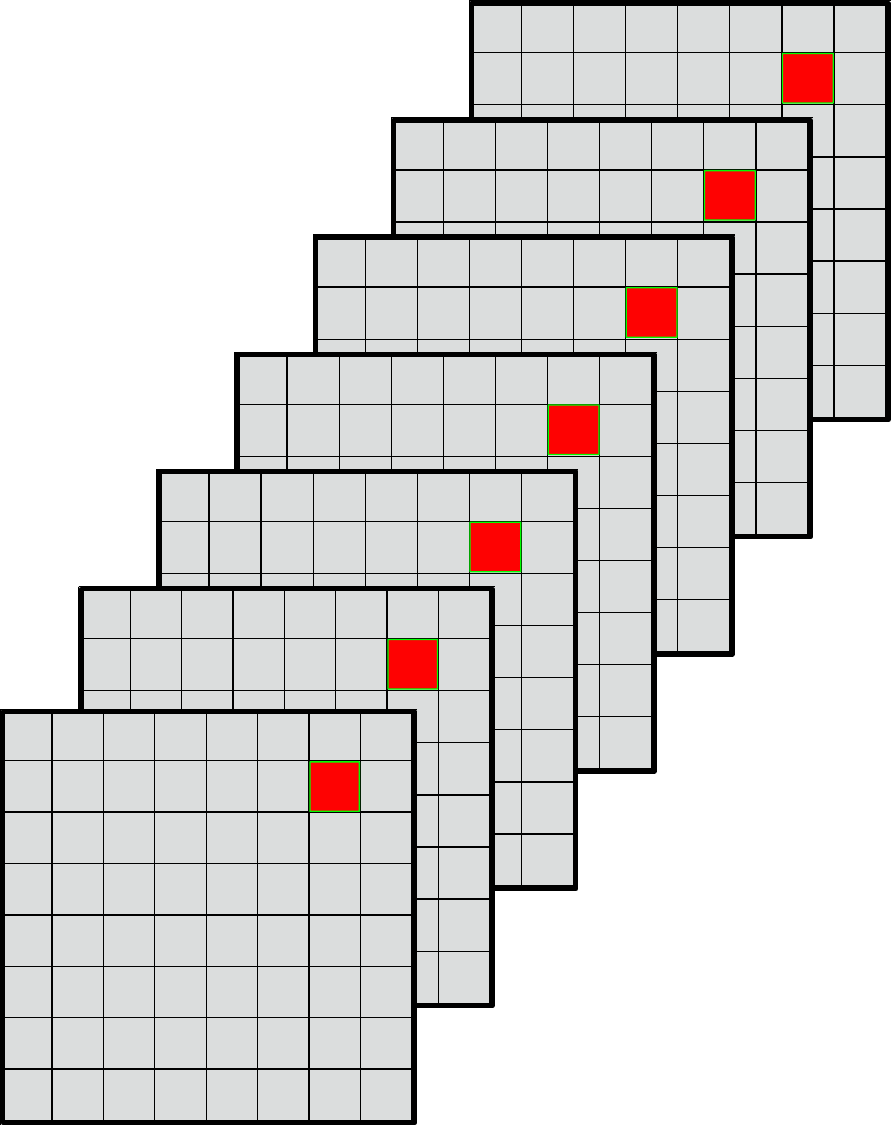}
  \includegraphics[width=.37\columnwidth]{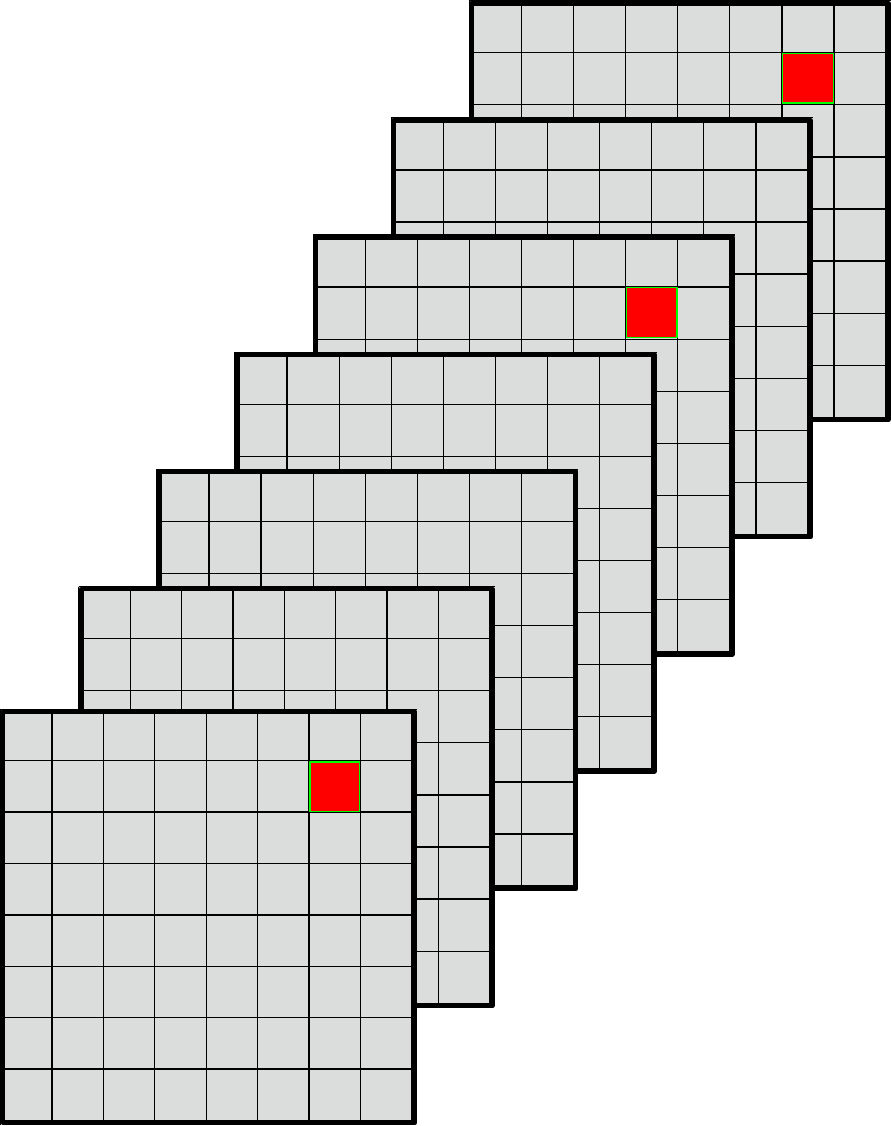}}
  \qquad
  \subfigure[Shattered glass\label{figure:shatter}]%
  {\includegraphics[width=.37\columnwidth]{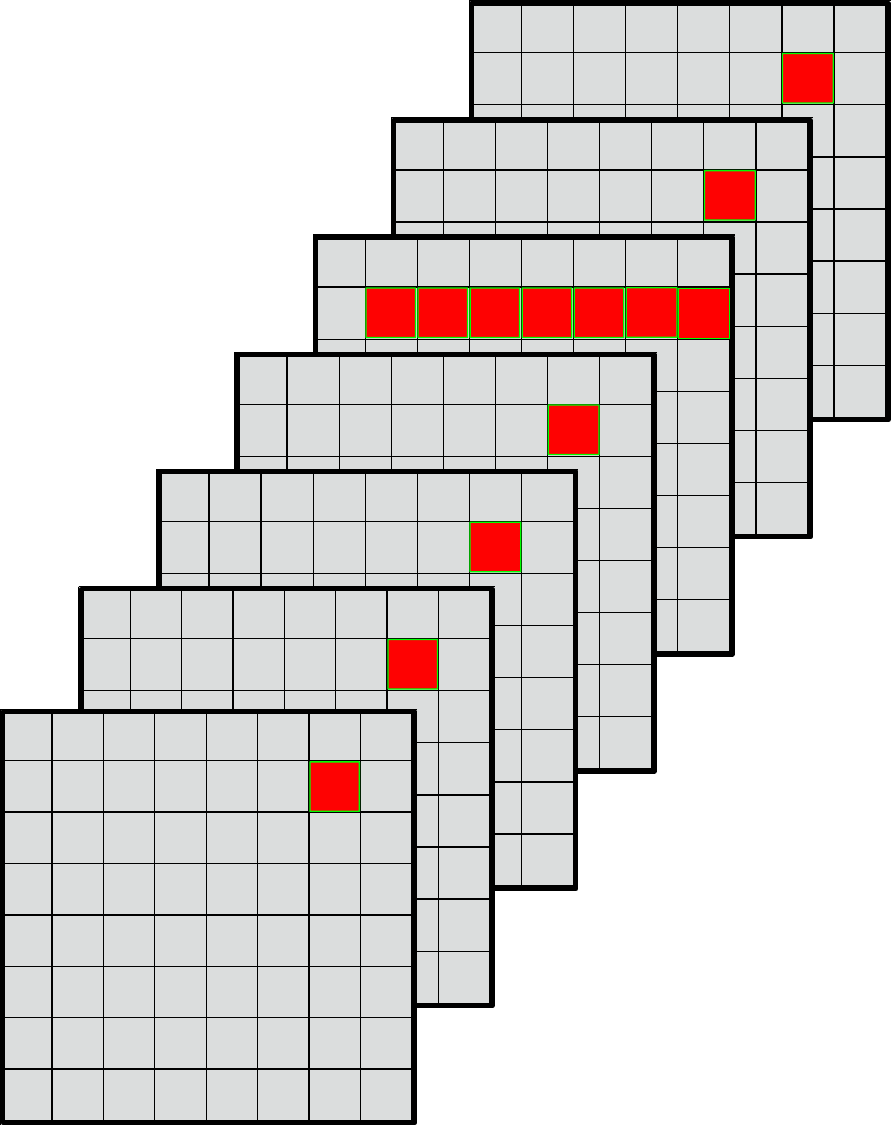}
  \includegraphics[width=.37\columnwidth]{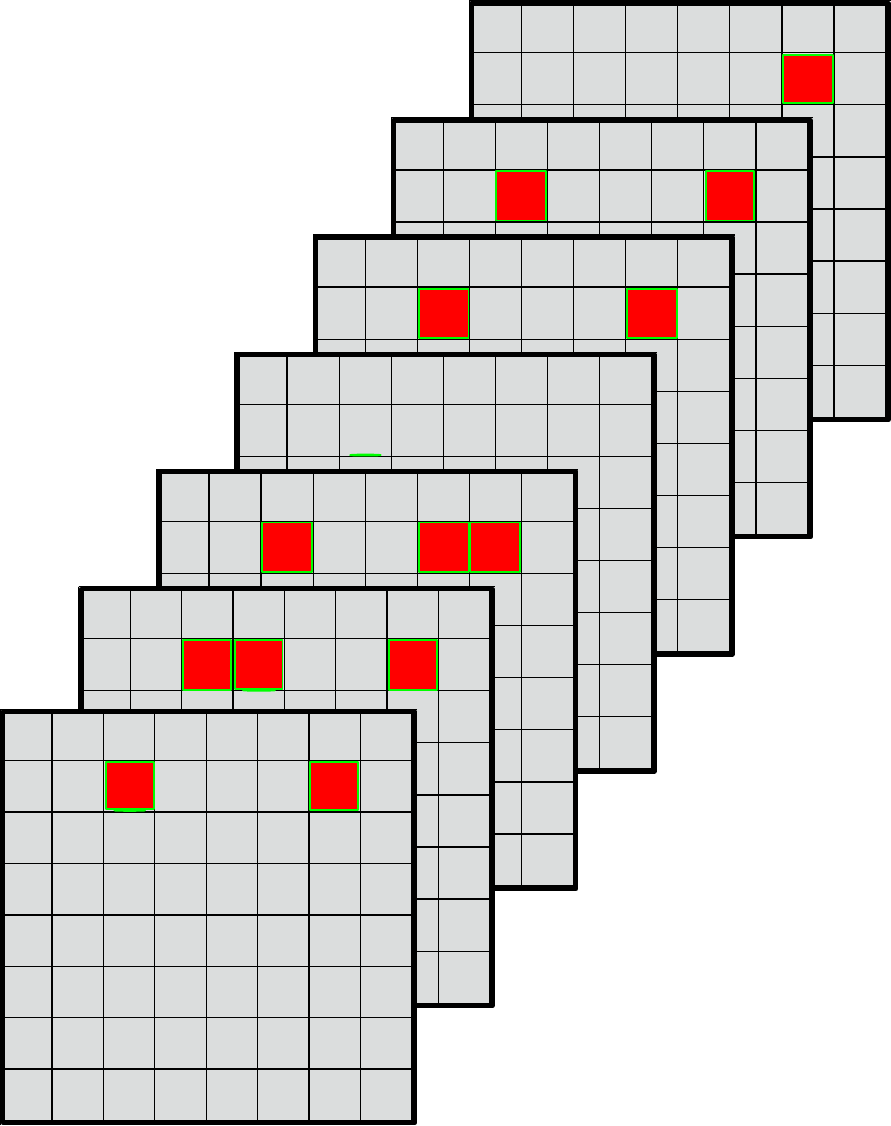}
  \includegraphics[width=.37\columnwidth]{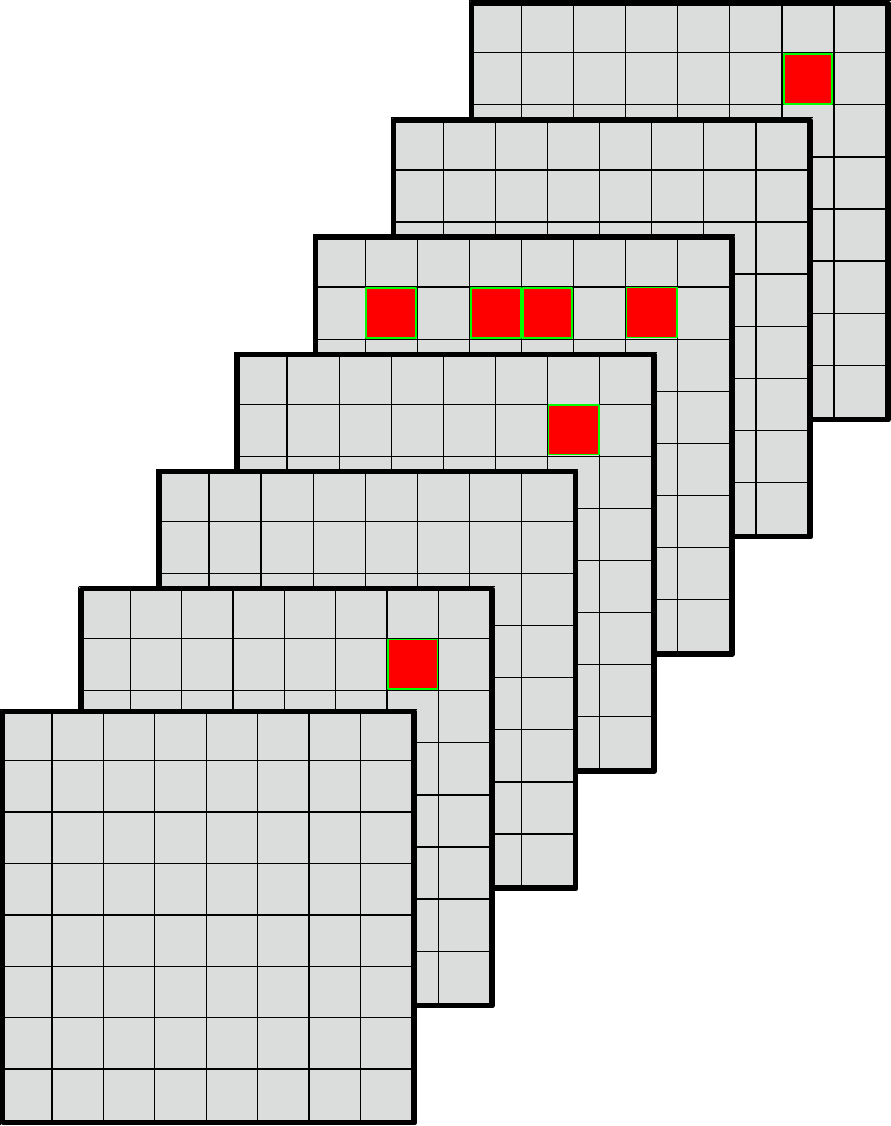}}
  \caption{Spatial distribution patterns (erroneous values are colored in red).  (a) \textit{Single point}: the fault causes the corruption of a single value of a single feature map. (b) \textit{Same row}: the fault causes the total or partial corruption of a row in a single feature map. (c) \textit{Bullet wake}: the fault corrupts the same location in all or multiple feature maps. (d) \textit{Shatter glass}: the fault causes the combination of the effects of \textit{same row} and \textit{bullet wake} patterns.\label{figure:spacial}}
 \end{figure*}

\subsubsection{Erroneous Values Cardinality}\label{sec:cardinality}

We first analyzed the number of erroneous values found in a corrupted output tensor; such a number is strongly dependent on the operator under analysis. 
For most considered operators the corrupted output tensor contains only one erroneous value in 85\% to 92\% of the experiments (depending on the operator) and two erroneous values in the remaining cases, while a higher cardinality is observed in less than 1\% of the cases. However, when considering the \texttt{Convolution} (\texttt{1}, \texttt{2}, or \texttt{3}) and \texttt{Batch normalization} operators, the frequency of having multiple corrupted tensors is much higher, ranging from 22\% to 57\%, as shown in Figure~\ref{figure:cardinality}.

This can be explained by considering that most operators are implemented as ``linear kernels'' where each thread computes one value of the output tensor without exploiting \ac{GPU} shared memory and with no cooperation among threads in the same block. The \texttt{Convolution} and \texttt{Batch normalization} operators are composed of several kernels based on \ac{GEMM} algorithm; thus, their implementations heavily exploit \ac{GPU} shared memory for threads cooperation.
The corruption of one of those threads can propagate to other threads and result in many errors in the output.

\subsubsection{Erroneous Values Domains}

As a second analysis, we studied how each erroneous value in the corrupted tensor deviates from the golden counterpart. In particular, we defined the domains the erroneous values fall into as follows:
\begin{itemize}
    \item \textit{Not a number}: the corrupted value is NaN,
    \item \textit{Zero}: the corrupted value is zero,
    \item \textit{Bitflip}: the corrupted value differs from the expected one by a single bit,
    \item \textit{[-1,1]}: the difference between the expected value and the actual one is between -1 and 1, and
    \item \textit{Random}: the value is corrupted in a completely random way, thus not falling into any of the above cases.
\end{itemize}

For most operators the great majority of faults cause an error in the \textit{[-1,1]} class (70\% up to 97\% of the experiments, depending on the considered operator) with the remaining cases falling in the \textit{Random} class. When dealing with the \texttt{Add} and \texttt{Exp} operators, we observe a great majority of cases falling in the \textit{Zero} class. Finally, for the \texttt{Div} operator the \textit{Random} and \textit{Bitflip} classes are predominant, while for the \texttt{Sigmoid} operator most of the erroneous values belong to \textit{Random} and \textit{[-1,1]}.
Figure~\ref{figure:domains} reports the observed distributions of the erroneous values for each considered operator.

\subsubsection{Spatial Distribution}
Finally, we analyzed the spatial distribution of the erroneous values within the corrupted output tensor. A first classification is between faults that produce one or more errors in a unique feature map and those whose effect spreads over multiple feature maps.

\noindent{\textbf{Single Feature Map:}} most of the faults that cause less than 16 erroneous values in the output tensor affect only a single feature map. This is due to the fact that the \ac{GPU} kernel is implemented to group threads working on the same feature map in a single block; thus such multiple errors may be due to the corruption of predicated instructions.
In this subfamily of spatial distributions we identify:
\begin{itemize}
    \item \textit{Single point}: a single value is corrupted (Figure~\ref{figure:smf_sp}),
    \item \textit{Same row}: multiple corrupted values lie in the same row, as in Figure~\ref{figure:sameRow}, and
    \item \textit{Random}: no regular pattern.
\end{itemize}

\noindent{\textbf{Multiple Feature Maps:}} most of the faults that cause more than 16 erroneous values in the output tensor spread the corrupted values among several feature maps; this is due to the fact that, as mentioned, such operators heavily exploit shared memories. In this subfamily of spatial distributions we identify:
\begin{itemize}
    \item \textit{Bullet wake}: the same location is corrupted in all (or in multiple) feature maps, as in Figure~\ref{figure:bullet},
    \item \textit{Shattered glass}: like one or more \textit{Bullet wake} errors, but in one or multiple feature maps the corruption spreads over a row (or part of the row), as in Figure~\ref{figure:shatter}, and
    \item \textit{Random}: no regular pattern.
\end{itemize}

Table~\ref{tab:spatialpatterns_distribution} reports the frequency of each spatial distribution pattern for each operator. It can be observed that, consistently with the considerations drawn in Subsection~\ref{sec:cardinality}, spatial distributions involving multiple feature maps only appear for the \texttt{Convolution} (\texttt{1}, \texttt{2}, or \texttt{3}) and the \texttt{Batch normalization} operators (the only operators exposing a cardinality of erroneous values greater than 2). It is also worth mentioning that, on average, random errors  have an incidence of 1.15\% and 5.2\% in single feature map and multiple features maps, respectively. To summarize, the great majority of the fault effects (93.65\% on average) actually cause a well-identifiable effect, that can be modeled and reproduced through an algorithmic description.

\begin{table}[t]
    \caption{Spatial patterns frequencies\label{tab:spatialpatterns_distribution}}
    \centering
\resizebox{\columnwidth}{!}{%
    \begin{tabular}{|l|r|r|r||r|r|r|}%r|}
    \hline
   \multirow{3}{*}{\textbf{Operator}}
    & \multicolumn{3}{c||}{\textbf{Same Feature Map}} & \multicolumn{3}{c|}{\textbf{Multiple Feature Maps}}\\
    \cline{2-7}
    
     & \multicolumn{1}{c|}{\textbf{Single}} & \multicolumn{1}{c|}{\textbf{Same}} & \multirow{ 2}{*}{\textbf{Random}} & \multicolumn{1}{c|}{\textbf{Bullet}} & \multicolumn{1}{c|}{\textbf{Shatter}}  & \multirow{ 2}{*}{\textbf{Random}}\\

    &\multicolumn{1}{c|}{\textbf{Point}} & \multicolumn{1}{c|}{\textbf{Row}} &  & \multicolumn{1}{c|}{\textbf{Wake}} & \multicolumn{1}{c|}{\textbf{Glass}}
    & \\
    \hline    \hline
    \texttt{Conv. 1} & 42.7\% & 18.7\% & %0.0 & 
    0.0\% & 20.6\% & 16.2\% & 1.8\%\\
    \hline
    \texttt{Conv. 2} & 40.9\% & 13.1\% & %0.0 & 
    0.1\% & 32.9\% & 11.5\% & 1.3\%\\
    \hline
    \texttt{Conv. 3} & 50.5\% & 12.4\% & %0.0 & 
    0.1\% & 25.4\% & 11.5\% & 0.1\%\\
    \hline
    \texttt{Add} & 90.3\% & 1.8\% & %0.0 & 
    0.8\% & 0.0\% & 0.0\% & 7.1\%\\
    \hline
    \texttt{Batch Norm.} & 77.8\% & 2.5\% & %0.0 & 
    1.1\% & 12.7\% & 1.1\% & 3.0\%\\
    \hline
    \texttt{Bias add} & 90.1\% & 1.2\% & %0.0 & 
    1.0\% & 0.2\% & 0.0\% & 7.5\%\\
    \hline
    \texttt{Div} & 84.9\% & 6.7\% & %0.0 & 
    8.4\% & 0.0\% & 0.0\% & 0.0\%\\
    \hline
    \texttt{Exp} & 91.9\% & 0.5\% & %0.0 & 
    0.0\% & 0.0\% & 0.0\% & 7.5\%\\
    \hline
    \texttt{ReLU} & 84.5\% & 1.5\% & %0.0 & 
    1.1\% & 0.0\% & 0.0\% & 10.3\%\\
    \hline
    \texttt{Mul} & 88.4\% & 0.3\% & %0.0 & 
    0.0\% & 0.0\% & 0.0\% & 11.3\%\\
    \hline
    \texttt{Sigmoid} & 89.7\% & 1.4\% & %0.0 & 
    0.0\% & 0.0\% & 0.0\% & 9.0\%\\
    \hline
\end{tabular}}
\end{table}

\subsection{Error Model Definition}
The detailed analysis carried out on the outcomes of the extensive \ac{FI} campaign is the solid basis on which we build the definition of functional error models for \ac{CNN} operators. More precisely, two are the key elements we identified to generalize from a corruption pattern to an error model:
\begin{inparaenum}[i)]
\item a statistical relevance of the effects, and
\item the possibility to specify an algorithm that produces the desired pattern. 
\end{inparaenum}
Based on these premises, the only spatial pattern that cannot be systematically and accurately reproduced through an algorithm is the \textit{Random} one, for which no error model has been defined. Since its statistical relevance is also modest (less than 6.35\%), the set of defined error models can be considered sound and complete. 

Furthermore, we identified a list of parameters for each model (reported in Table~\ref{table:parameters}), to make it general and exploitable in an automatic simulation engine.
In addition to the listed parameters, when generating a \textit{Same row} error a number of points in the row is randomly left unaltered. Similarly for the \textit{Bullet wake} and \textit{Shattered glass} errors, a number of feature maps is randomly left unaltered.

The resulting set of error models well covers the effects caused by the faults and can be conveniently exploited in an error simulation campaign, for an early and accurate analysis of the reliability against \ac{SEU} faults of a \ac{CNN}-based application executed on a \ac{GPU}.

\begin{table}[t]
    \centering
    \caption{Error model parameters\label{table:parameters}}
\resizebox{\columnwidth}{!}{
    \setlength\tabcolsep{3pt}
    \begin{tabular}{|l|l|}%p{6cm}}
    \hline
    \textbf{Error model} &  \textbf{Parameters}\\ 
    \hline
    \textit{Single point} & x and y coords. of the point\\ \hline 
    \textit{Same row} & x and y coords. of start and end point of the row \\ \hline 
    \multirow{2}{*}{\textit{Bullet wake}} & x and y coords. of the enter point, \\ & first feat. map, last feat. map \\ \hline 
    \multirow{2}{*}{\textit{Shattered glass}} & x and y coords. of the enter point, \\  & first feat. map, last feat. map, shattered feat. map\\ \hline 
    \end{tabular}
    }
\end{table}

\bigskip

The outcome of the adopted process is general with respect to the \ac{CNN} operators and the target platform. However, the process is designed to fine tune the fault injection campaign and the error modeling activity, for an accurate outcome in relation to the case study under consideration. Therefore, when a new \ac{CNN} is taken into account, it might require a re-execution of these activities, to guarantee an accurate and updated error model repository.

\section{Error Simulation Results}
\label{sec:simulation}
The defined error models have been integrated in \ac{ourtool} error simulation engine. 
A final campaign has been prepared to analyze the ability of the proposed error simulator in managing a complete \ac{CNN}. Moreover, we compared our proposal against the state-of-the-art fault injection and error simulation tools, namely SASSIFI and TensorFI, respectively.

All the experiments and comparisons have been performed on a machine equipped with an Intel Core\textsuperscript{\textregistered} i7-4870HQ as CPU and a NVIDIA GeForce GT 750M as GPU, running on Ubuntu 18.04 LTS.

\subsection{\ac{ourtool} vs. state-of-the-art \ac{FI}}
SASSIFI offers fault injection facilities, therefore the comparison between this state-of-the-art solution and our proposal considers both accuracy and performance measured in the execution time.
For this experimental comparison we considered YOLO V3 \ac{CNN}. We re-used the same fault list of the previous section, consisting of 360,000 random fault injections.

First of all we have to mention that we could not compare against a \textit{pure} SASSIFI experiment where the entire YOLO \ac{CNN}, implemented in Caffe, is loaded onto the \ac{FI} environment due to scalability issues. Indeed, a single YOLO run on SASSIFI took about 15 minutes, which would have led to about 10 years for running 360,000 experiments. Therefore we considered a \textit{hybrid} SASSIFI configuration, coupled with TensorFlow; we limited the execution through SASSIFI only for the operator being corrupted by the fault, while the rest of the \ac{CNN} is executed with TensorFlow at full speed. This workaround allows the extensive \ac{FI} campaign to be executed using state-of-the-art tools.
More in details, the 360,000 injected faults allowed to collect 137,004 corrupted faulty outputs, in about 92 hours. To perform the same 137,004 relevant error simulation experiments \ac{ourtool} took about 15 hours, corresponding to a 6x speedup with respect to the hybrid SASSIFI configuration. However, the speed up is beneficial but not paramount, considering the rapid evolution of new tools (in the future we will investigate the integration of NVBitFI), and it has been analyzed/compared to evaluate the efficiency of our solution.

The injection of the same error in the same operator for both environments allows the accuracy comparison.
In particular, we considered the output of SASSIFI as ground truth (being it an accurate architecture-level fault injection tool) and we compared the effect of the fault (in SASSIFI) and of the error (in \ac{ourtool}) on the final classification produced by YOLO, to evaluate if both engines produce the same type of output or not (usable/unusable). 
YOLO performs an object detection task by computing bounding boxes around each identified object in the analyzed picture. In the discussed experiment, we were interested in studying whether the \ac{CNN} was able to correctly detect the objects despite their actual position. For this reason, we defined an output classifier that tags the output of \ac{CNN} as usable only if it contains the same list of detected objects as the golden counterpart.
Collected data show an average $98.72\%$ accuracy w.r.t. SASSIFI for the proposed \ac{ourtool}.

\subsection{\ac{ourtool} vs. state-of-the-art \ac{FES}}
TensorFI offers \ac{FES} facilities, based on a limited fault model with respect to the one we defined and use in the proposed \ac{FES} activity. 
We did not perform any qualitative comparison between our results and the ones obtained by TensorFI because the error models implemented by TensorFI, basically bit-flips or single value corruptions, are a reduced subset of the ones we identified and implemented. 
The comparison against the state-of-the-art solution is therefore carried out considering only the performance of the approaches, given the previous discussion on the cardinality and adherence of the fault models.

Indeed,  we were not able to execute YOLO with TensorFI, for two reasons mainly; TensorFI does not support all the operators, e.g., Leaky ReLu and Batch normalization, and the tensor data format (Batch, Height, Width, Channel -- BHWC vs. Batch, Channel, Height, Width -- BCHW) employed by YOLO and it would have required too long execution times, according to~\cite{chen2020}.
For this reason, we here introduce two simpler \acp{CNN} to carry out the performance comparison: LeNet-5~\cite{lecun1989backpropagation} and CIFAR10~\cite{CIFAR10I1:online} 
and gather a trend in the achievable speedup. LeNet-5 is used for hand-written digits classification for the MNIST dataset~\cite{MNISThan12:online}, 
able to achieve a $99.05\%$ accuracy. It is a simple network composed of two convolutional layers and three dense layers. CIFAR10 performs object classification for the CIFAR10 dataset~\cite{CIFAR10I1:online}. 
In particular, we employed the Keras implementation\footnote{Keras: Deep Learning for humans, \url{https://github.com/awslabs/keras-apache-mxnet}, (Accessed on 12/01/2021).}, that achieves an accuracy equal to the $78\%$.

For this test, we defined a campaign of 10,000 random error simulations.
The execution times (in minutes) of TensorFI and \ac{ourtool} for the various campaigns are reported in Table~\ref{tab:timeES}. 
We also report the time for the nominal execution of the \ac{CNN} to allow the reader to get an idea of the minimal overhead introduced by our error injection mechanism.

As it can be noticed, \ac{ourtool} is able to execute the experiment on YOLO in a reasonable time, i.e. with a 2.5x slow down.
Moreover, \ac{ourtool} considerably outperforms the state-of-the-art solution on these small \acp{CNN}, with a speedup of about 44x and 63x. The motivation is that, as commented in~\cite{chen2020}, TensorFI re-defines \ac{CNN} operators in Python to perform the injection and such implementation is not accelerated onto \ac{GPU} as the original TensorFlow counterpart.
At the opposite, \ac{ourtool} is exclusively based on TensorFlow public \ac{API}, thus benefiting from the \ac{GPU} acceleration of the operators. Moreover, the defined input caching strategy allows for an additional performance improvement. 
As a final remark, the time required for the instrumentation of the \ac{CNN} to run the error simulations is about 80s, which is negligible for both approaches.

\begin{table}[tbp]
\centering
\caption{Execution times for \ac{FES} (10,000 error simulation experiments)}\label{tab:timeES}
\begin{tabular}{|l|rrr|}
\hline
& \multicolumn{3}{c|}{\ac{CNN}}\\
& \multicolumn{1}{c}{YOLO} & \multicolumn{1}{c}{LeNet-5} & \multicolumn{1}{c|}{CIFAR10} \\ \hline
%SASSIFI            & 883h                     & 53h                         & 84h                         \\
TensorFI & - & 18min & 40min\\
\ac{ourtool} & 25min & 0.41min & 0.63min\\
Plain execution (10,000 runs) & 10min  & 0.27min& 0.41min\\
\hline
\end{tabular}
\end{table}

\section{Advantages of the proposed approach}
\label{sec:adv}
We here recap the several advantages offered by the proposed cross-layer framework with respect to the state-of-the-art \ac{FI} and \ac{FES} environments.

\noindent\textbf{Accuracy.}
The exploitation of error models directly extracted from \ac{FI} experiments provide an overall  accuracy comparable with the one achieved by an \ac{FI} tool.

\noindent\textbf{Speed.}
The adoption of the \ac{FES} paradigm not affected by fault activation issues allows for the framework to be much faster than the classical \ac{FI} tools in achieving the desired amount of corrupted application outputs.

\noindent\textbf{Increased productivity.}
The integration of the reliability analysis tool within the standard \ac{ML} development framework used to design and train \acp{CNN} avoids the extra time and effort needed for porting the \ac{CNN} to different tools, each one performing a portion of the overall analysis. The same unified environment is now used to analyze and design \acp{CNN} from all points of view.

\noindent\textbf{Ease of use.} The framework is almost fully automated, thus easing the designer's activity. Indeed, being the number of \ac{CNN} operators limited, SASSIFI may be dismissed after analyzing a few \ac{CNN} applications.

\noindent\textbf{Flexibility and customizability.} 
The implementation in a scripting language gives a high flexibility, extensibility and customizability of the source code to meet the needs of each \ac{CNN} being analyzed at a very low designer's effort, as, for instance, to specify the output classification function.

\noindent\textbf{Portability and minimal intrusiveness.}
The instrumentation performed by \ac{ourtool} on the \ac{CNN} graph to include the saboteurs is minimal and automated. No change has been introduced either into the TensorFlow library or in the operators used in the \ac{CNN} graph. This allows the framework to be portable among different machines and to obtain a quite limited performance degradation w.r.t. nominal execution, as shown in the experimental results. 

Finally, it is worth commenting that, even if the tool has been implemented in TensorFlow, it can be potentially ported to any other \ac{ML} framework having a similar dataflow graph model and \ac{API} (e.g. PyTorch or Keras).

\section{Conclusions}
\label{sec:lausdeo}

We presented \ac{ourtool}, a cross-layer framework for the reliability analysis of \acp{CNN} that combines the accuracy of architecture-level fault injection with the ease of use, controllability, flexibility and speed of error simulation. 
We compared our methodology against the state-of-the-art SASSIFI fault injection environment and TensorFI functional error simulator, highlighting how our methodology achieves about 99\% accuracy in terms of the ability of reproducing the effects of faults on the final \ac{CNN} output with about 6x speedup w.r.t. SASSIFI, and a speedup ranging from 44x up to 63x w.r.t. TensorFI.

Future work is devoted to extend the framework to handle other platforms and ML algorithms, always keeping flexibility and extensibility in mind.


\begin{IEEEbiography}[{\includegraphics[width=1in,height=1.25in,clip,keepaspectratio]{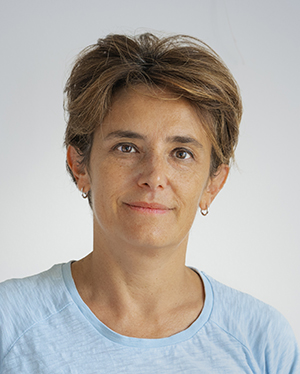}}]
{Cristiana Bolchini} is a Professor at Politecnico di Milano, where she received a Ph.D. in Automation and Computer Engineering in 1997. Her research interests cover the areas of methodologies for the design and analysis of digital systems with a specific focus on dependability and self-awareness for heterogeneous system architectures. She has co-authored more than 130 papers in peer-reviewed international journals and conferences. In 2020, she served as the Technical Programme Chair of the conference IEEE/ACM Design Automation and Test in Europe.
\end{IEEEbiography}

\begin{IEEEbiography}[{\includegraphics[width=1in,height=1.25in,clip,keepaspectratio]{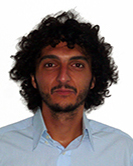}}]
{Luca Cassano} is an Assistant Professor at Politecnico di Milano, Italy. He received the B.S., M.Sc. and Ph.D. degrees in Computer Engineering from the University of Pisa, Italy. His research activity focuses on the definition of innovative techniques for fault simulation, testing, untestability analysis, diagnosis, and verification of fault tolerant and secure digital circuits and systems. With his Ph.D. thesis, titled ``Analysis and Test of the Effects of Single Event Upsets Affecting the Configuration Memory of SRAM-based FPGAs'', he won the European semifinals of the 2014 TTTC's E. J. McCluskey Doctoral Thesis Award and he classified as runner-up at the world finals.
\end{IEEEbiography}

\begin{IEEEbiography}[{\includegraphics[width=1.in,height=1.25in,clip,keepaspectratio]{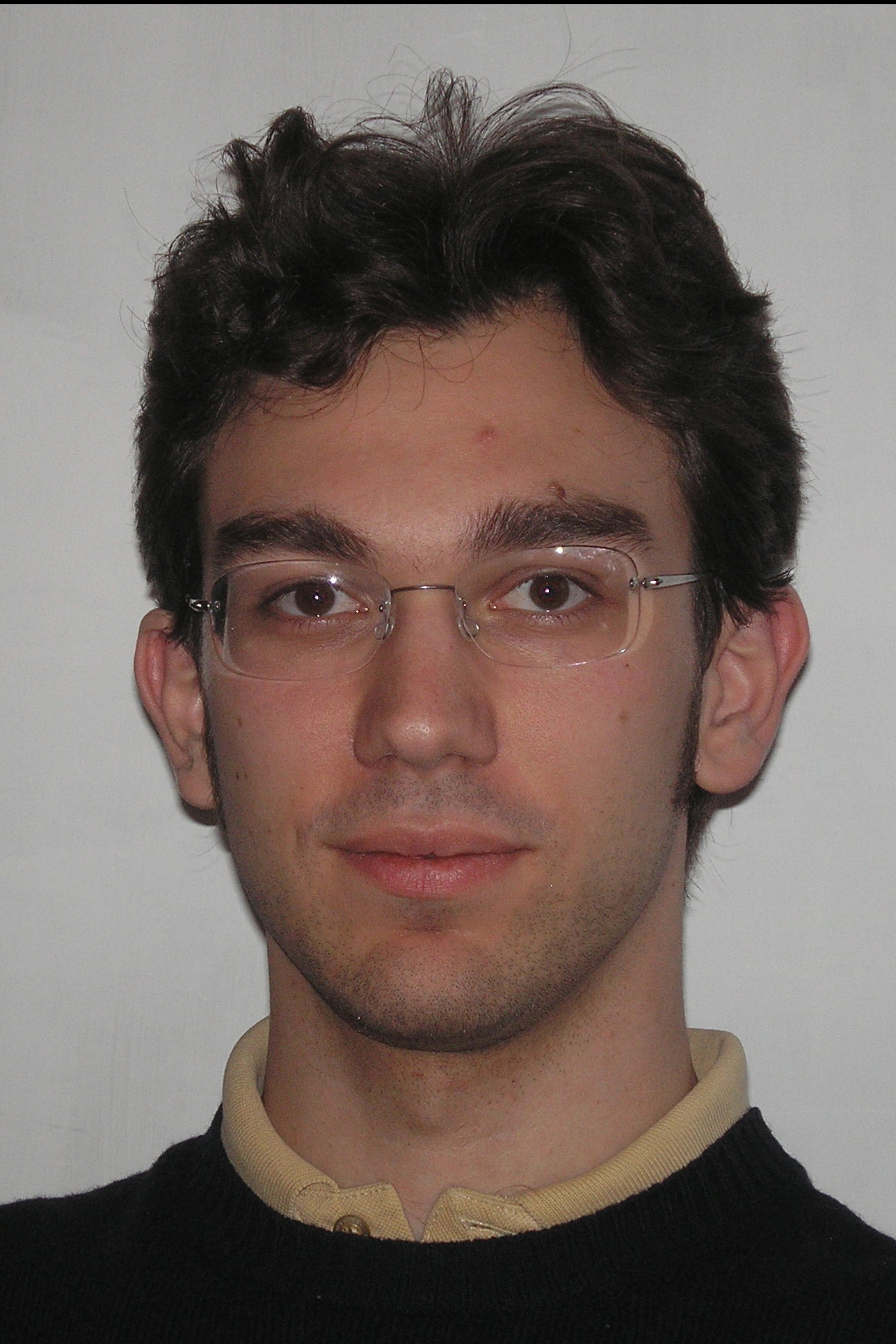}}]
{Antonio Miele} is an Associate Professor at Politecnico di Milano. He holds a M.Sc. in Computer Engineering from Politecnico di Milano and a M.Sc. in Computer Science from the University of Illinois at Chicago. In 2010 he received a Ph.D. degree in Information Technology from Politecnico di Milano. His main research interests are related to design methodologies for embedded systems, in particular fault tolerance and reliability issues, runtime resource management in heterogeneous multi-/many-core systems and FPGA-based systems design.
He is co-author of more than 80 peer-reviewed publications. 
\end{IEEEbiography}

\begin{IEEEbiography}[{\includegraphics[width=1in,height=1.25in,clip,keepaspectratio]{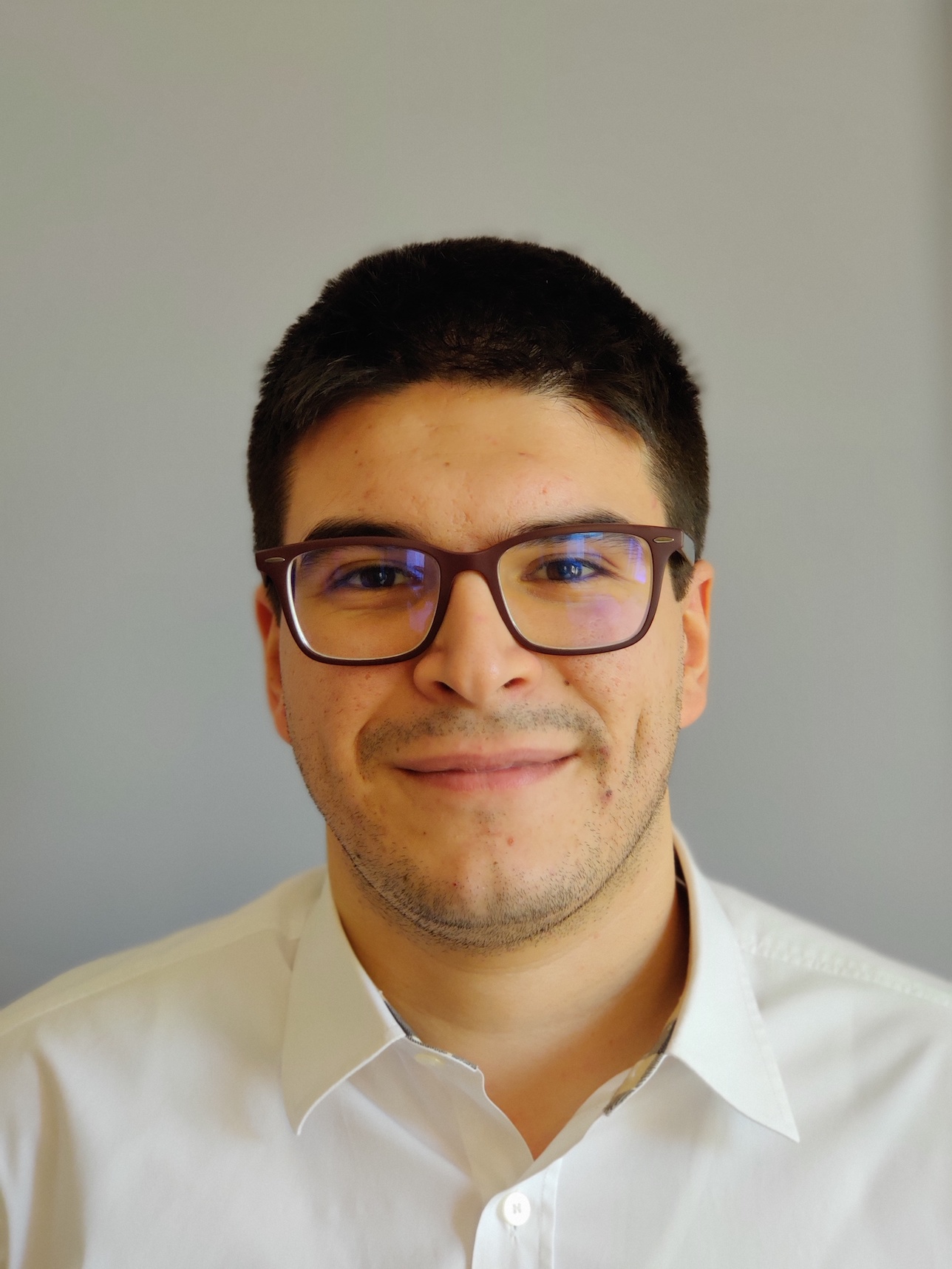}}]
{Alessandro Toschi} is a Computer Science Engineer and holds an M.Sc. in Computer Science Engineering from Politecnico di Milano in April 2020. During his master, he attended a double degree program between Politecnico di Milano and Shanghai Jiao Tong University, achieving an M.Sc. in Computer Science and Technology in April 2021. His master thesis regarded the reliability analysis of CNNs executed on GPUs, while his interests are in the fields of Heterogeneous Computing Systems and High Performance Computing. He is currently working as a mobile engineer, implementing photo/video editing algorithms on mobile GPUs.
\end{IEEEbiography}

\end{document}